\documentclass[useAMS,usenatbib]{mn2e}
\usepackage{amsmath,amssymb,graphicx}

\newcommand{\Msun}{\ensuremath{\,{M}_\odot}}                      
\newcommand{\Rsun}{\ensuremath{\,{R}_\odot}}                      
\newcommand{\Teff}{\ensuremath{T_{\rm eff}}}                      
\newcommand{\Mjup}{\ensuremath{\,{M}_{\rm Jup}}}                  
\newcommand{\Rjup}{\ensuremath{\,{R}_{\rm Jup}}}                  
\newcommand{\Teq}{\ensuremath{T_{\rm eq}^{\,\prime}}}             
\newcommand{\safronov}{\ensuremath{\Theta}}                       
\newcommand{\kms}{\,km\,s$^{-1}$}                                 
\newcommand{\mss}{\,m\,s$^{-2}$}                                  
\newcommand{\pjup}{\ensuremath{\,\rho_{\rm Jup}}}                 
\newcommand{\psun}{\ensuremath{\,\rho_\odot}}                     
\newcommand{\mc}[1]{\multicolumn{2}{c}{#1}}
\newcommand{\mcc}[1]{\multicolumn{3}{c}{#1}}

\newcommand{\erc}[3]{\mc{\ensuremath{#1^{+#2}_{-#3}}}}

\title[Physical properties of Qatar-2b]
{Physical properties, starspot activity, orbital obliquity, and
transmission spectrum of the Qatar-2 planetary system from
multi-colour photometry\thanks{Based on data collected with GROND
at the MPG/ESO 2.2-m telescope, BUSCA at the CAHA 2.2-m telescope,
BFOSC at the Cassini 1.52-m telescope, and DLR-MKIII camera at the
CAHA 1.23-m telescope.}}

\author[L.~Mancini et al.]
{\parbox{\textwidth}{L.~Mancini$^{1}$\thanks{E-mail:
\texttt{mancini@mpia.de}}, %
        J.\ Southworth\,$^{2}$,
        S.\ Ciceri\,$^{1}$,
        J.\ Tregloan-Reed\,$^{2}$,
        I.\ Crossfield\,$^{1}$,
        N.\ Nikolov\,$^{3}$,
        I.\ Bruni\,$^{4}$
        R.\ Zambelli\,$^{5}$ and
        Th.\ Henning\,$^{1}$} \vspace{0.4cm}\\
\parbox{\textwidth}{
        $^{1}$\,Max Planck Institute for Astronomy, K\"onigstuhl 17, 69117 Heidelberg, Germany \\
        $^{2}$\,Astrophysics Group, Keele University, Staffordshire, ST5 5BG, UK \\
        $^{3}$\,Astrophysics Group, University of Exeter, Stocker Road, EX4 4QL, Exeter, UK \\
        $^{4}$\,INAF -- Osservatorio Astronomico di Bologna, Via Ranzani 1, 40127 -- Bologna, Italy \\
        $^{5}$\,Societ\`{a} Astronomica Lunae, 19030 Castelnuovo Magra (La Spezia), Italy
}}

\begin{document} %
\maketitle

\begin{abstract}
We present seventeen high-precision light curves of five transits
of the planet Qatar-2\,b, obtained from four defocussed 2m-class
telescopes. Three of the transits were observed simultaneously in
the SDSS $g^\prime r^\prime i^\prime z^\prime$ passbands using the
seven-beam GROND imager on the MPG/ESO 2.2-m telescope. A fourth
was observed simultaneously in Gunn $grz$ using the CAHA 2.2-m
telescope with BUSCA, and in $r$ using the Cassini 1.52-m
telescope. Every light curve shows small anomalies due to the
passage of the planetary shadow over a cool spot on the surface of
the host star. We fit the light curves with the {\sc prism+gemc}
model to obtain the photometric parameters of the system and the
position, size and contrast of each spot. We use these photometric
parameters and published spectroscopic measurements to obtain the
physical properties of the system to high precision, finding a
larger radius and lower density for both star and planet than
previously thought. By tracking the change in position of one
starspot between two transit observations we measure the orbital
obliquity of Qatar-2\,b to be $\lambda=4.3^{\circ} \pm
4.5^{\circ}$, strongly indicating an alignment of the stellar spin
with the orbit of the planet. We calculate the rotation period and
velocity of the cool host star to be $11.4 \pm 0.5$\,d and $3.28
\pm 0.13$\kms\ at a colatitude of $74^{\circ}$. We assemble the
planet's transmission spectrum over the $386$--$976$\,nm
wavelength range and search for variations of the measured radius
of Qatar-2\,b as a function of wavelength. Our analysis highlights
a possible H$_{2}$/He Rayleigh scattering in the blue.
\end{abstract}


\section{Introduction}
\label{sec:1}

Transiting extrasolar planets (TEPs) are the most interesting
exoplanets to study as it is possible to deduce their physical
properties to high precision. High-quality photometric
observations of TEPs are a vital component of such work, as
they strongly constrain the density of the host star
\citep{seager2003}. They also allow searches for transit timing
variations (TTVs; e.g.\ \citealp{holman2010}), which can be used
to measure the masses of the transiting planets or show the
presence of non-transiting objects \citep{nesvorny13}, and for
variations of the planetary radius with wavelength which trace
opacity variations in the planet's atmosphere.

Since 2008 we have been photometrically following up known TEP
systems from both hemispheres. The aim of this project is to
obtain high-precision differential photometry of complete transit
events, which can be used to refine the measured physical
properties of the planets and parent stars
\citep[e.g.][]{southworth2010al,southworth2011al,southworth2012al1,southworth2012al2,southworth2012al3,southworth2013al},
search for opacity-induced planetary radius variations
\citep[e.g.][]{mancini2013a,mancini2013b,mancini2013c,mancini2014,nikolov2013},
and investigate starspot crossing events
\citep{mohler2013,ciceri2013}. Our observations are performed
using medium-class defocussed telescopes, some of which are
equipped with multi-band imaging instruments.

In this work we present extensive new follow-up photometry of
Qatar-2, the second planetary system discovered by the Qatar
Exoplanet Survey (QES) \citep{bryan2012}. This system comprises
Qatar-2\,A, a moderately bright ($V=13.3$\,mag) K dwarf, which is
orbited by Qatar-2\,b, a 2.5\Mjup\ planet on a 1.34\,d
period\footnote{The discovery paper also reported the possible
existence of a second planet, Qatar-2\,c, in a $\sim$1\,yr orbit.
However, a recent Erratum \citep{bryan2014} has shown, using
additional radial velocity measurements, than there was an error
in the barycentric correction and that the outer planet was just a
detection of Earth's orbital motion.}. The late spectral type of
the host star means that the transits due to Qatar-2\,b are deep
and may contain starspot crossing events
\citep[e.g.][]{sanchis2011a,sanchis2011b,tregloan2013}.

We report observations of three transits simultaneously observed
in four optical passbands using the ``Gamma Ray Burst Optical and
Near-Infrared Detector'' (GROND) at the MPG/ESO 2.2-m telescope,
one transit simultaneously observed in three optical passbands
with the ``Bonn University Simultaneous CAmera'' (BUSCA) at the
CAHA 2.2-m telescope, one transit with the Cassini 1.5-m
telescope, one transit with the CAHA 1.23-m telescope and three
further transits observed with a 25-cm telescope. We use these new
light curves to refine the physical properties of the system and
attempt to probe the atmospheric composition of Qatar-2\,b at
optical wavelengths ($386$--$976$\,nm).

\section{Observational strategies to observe planetary transits}
\label{sec:2}

\begin{figure*}
\centering
\includegraphics[width=14.cm]{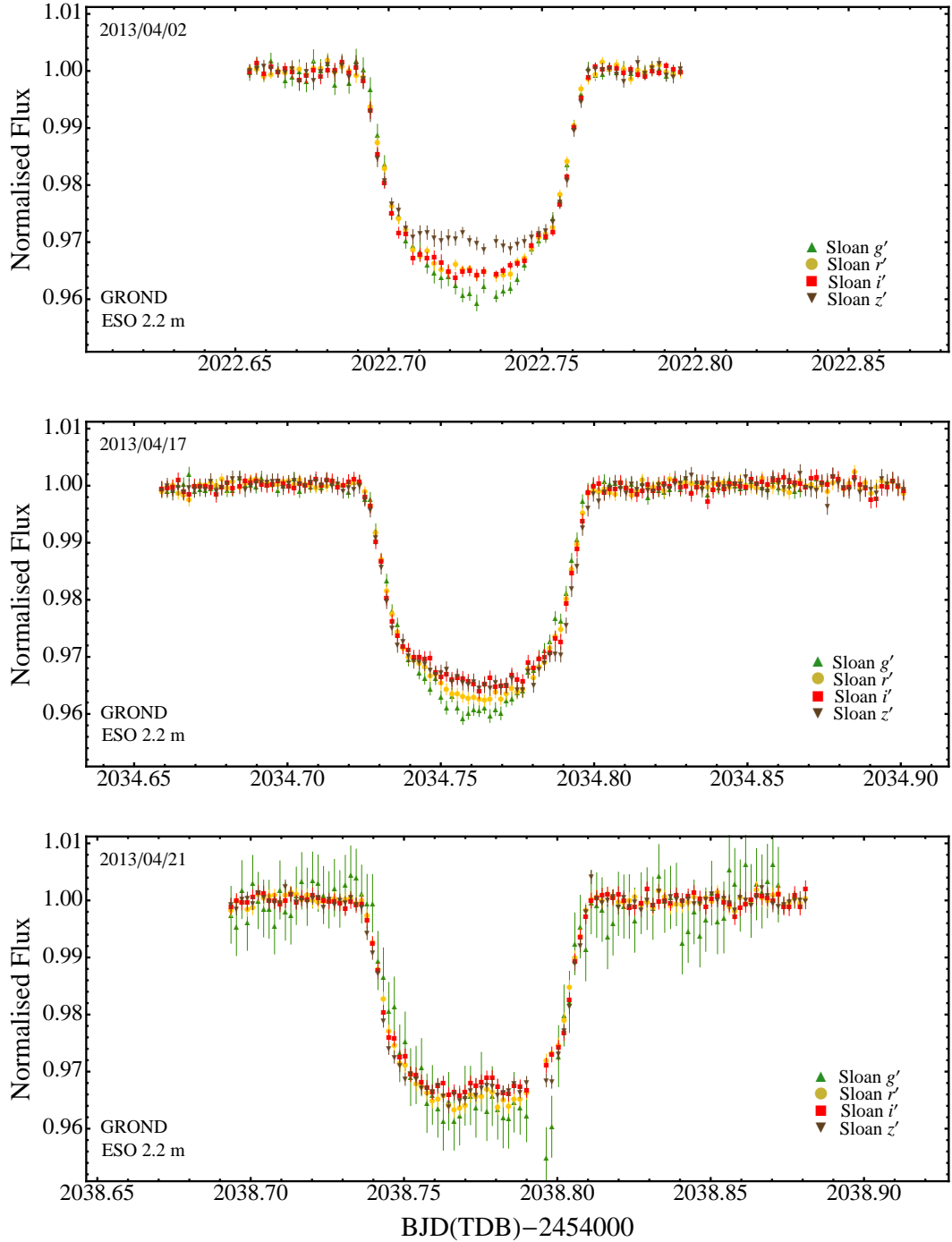}
\caption{Light curves of three transits of Qatar-2\,b observed
simultaneously in four optical bands with GROND, ordered according
to date.} \label{Fig:01}
\end{figure*}

\begin{figure*}
\centering
\includegraphics[width=14.cm]{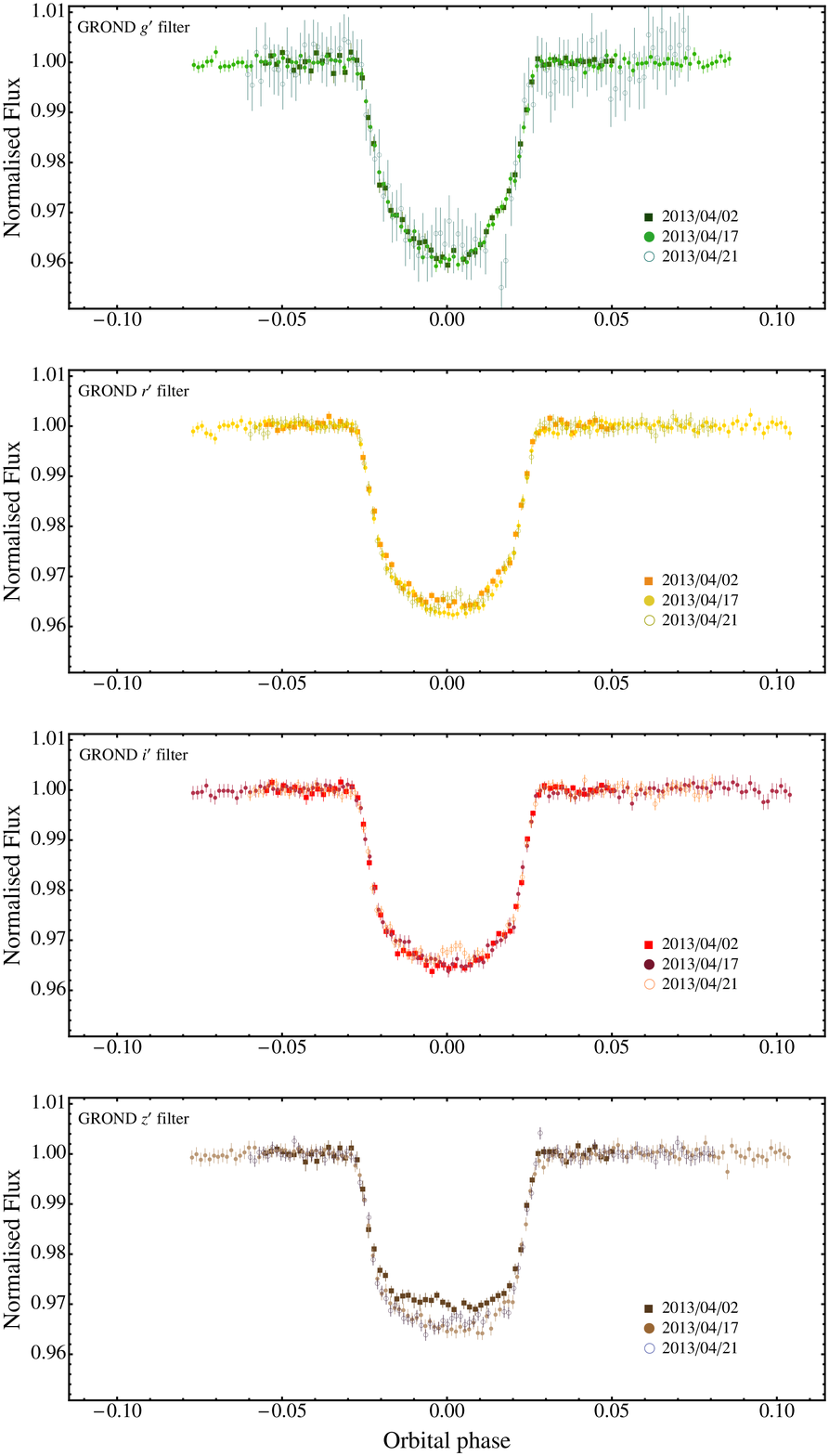}
\caption{Phased light curves of the three transits of Qatar-2\,b
observed with GROND, ordered according to the filter used. This
highlights the anomalies between the three transits in each
colour.} \label{Fig:02}
\end{figure*}

In this section we describe the methodologies used to obtain
accurate photometric observations of transiting-planet events and
get reliable physical information on the planetary system.

\begin{figure*}
\centering
\includegraphics[width=12.cm]{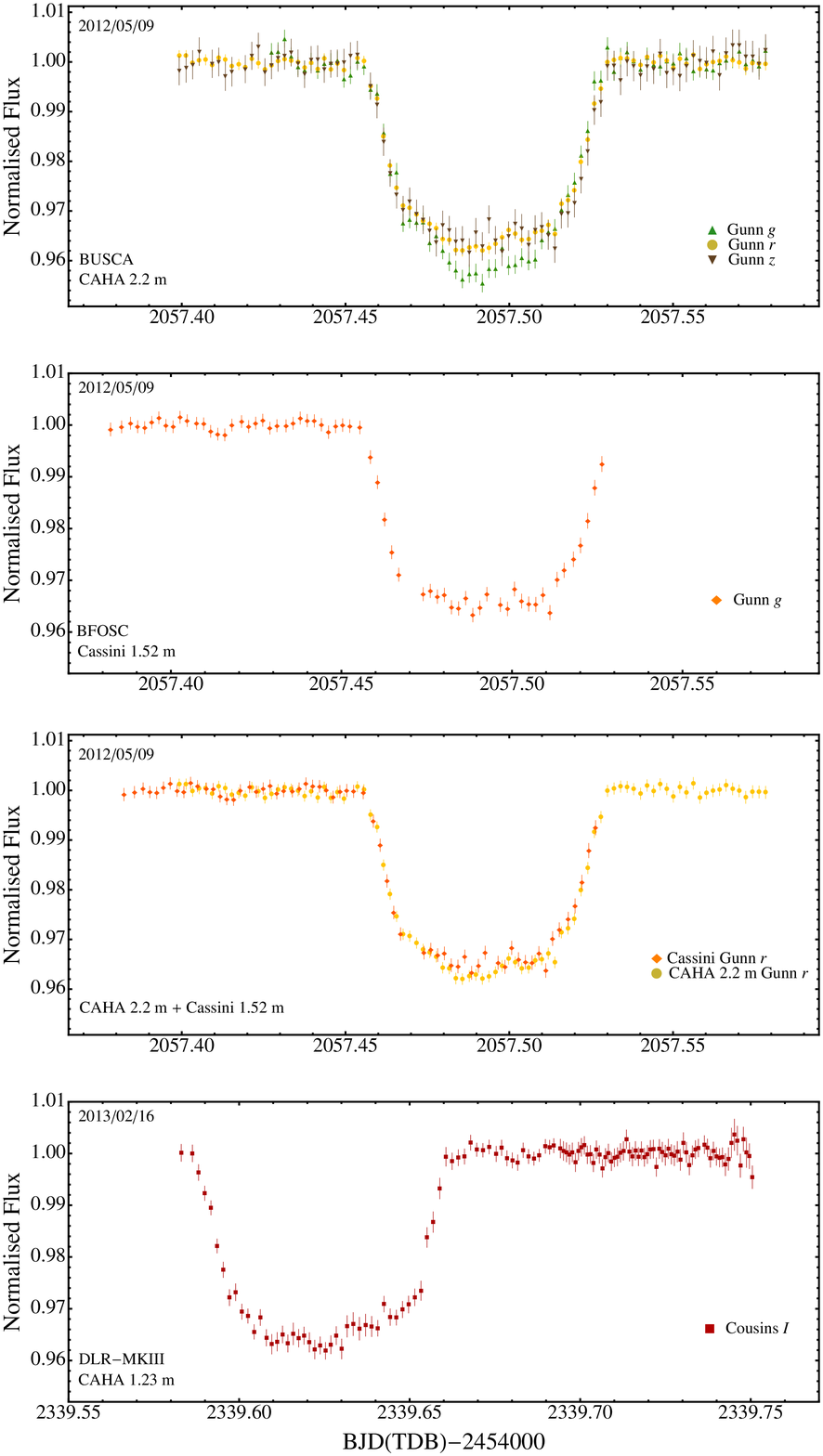}
\caption{Light curves of two transits of Qatar-2\,b observed with
three telescopes, shown in date order. \textit{Top panel}: light
curves of one transit of Qatar-2\,b observed simultaneously in
three optical bands with BUSCA at the CAHA 2.2\,m telescope. The
anomalies on the light curves are interpreted as the occultation
of a starspot by the planet. \textit{Second panel}: light curve
obtained with the Cassini 1.52\,m telescope using a Gunn $r$
filter. This is the same transit observed with BUSCA.
\textit{Third panel}: CAHA 2.2\,m and Cassini 1.5\,m Gunn-$r$
light curves. \textit{Bottom panel}: light curve obtained with the
CAHA 1.23\,m telescope using a Cousin $I$ filter.} \label{Fig:03}
\end{figure*}

\begin{table*}
\centering \tiny %
\caption{Details of the transit observations presented in this
work. $N_{\rm obs}$ is the number of observations, $T_{\rm exp}$
is the exposure time, $T_{\rm obs}$ is the observational cadence,
and `Moon illum.' is the fractional illumination of the Moon at
the midpoint of the transit. The aperture sizes are the radii of
the software apertures for the star, inner sky and outer sky,
respectively. Scatter is the r.m.s.\ scatter of the data versus a
fitted model. $\beta$ is the ratio between the noise levels due to
Poisson noise and to combined Poisson and red noise.}
\label{Table:1}%
\begin{tabular}{lcccrrccccccc}
\hline
Telescope & Date of   & Start time & End time & $N_{\rm obs}$ & $T_{\rm exp}$ & $T_{\rm obs}$ & Filter & Airmass & Moon   & Aperture   & Scatter & $\beta$\\
          & first obs. & (UT)       & (UT)     &                    & (s)           & (s)           &        &    & illum. & radii (px) & (mmag) \\
\hline
ESO~2.2\,m \#1  & 2012 04 02 & 03:34 & 06:56 & 61 & 120 & 200 & Sloan $g^{\prime}$ & $1.25 \rightarrow 1.08 \rightarrow 1.14$ & $83\%$ & 23, 45, 75 & 1.22 & 1.26 \\
ESO~2.2\,m \#1  & 2012 04 02 & 03:34 & 06:56 & 61 & 120 & 200 & Sloan $r^{\prime}$ & $1.25 \rightarrow 1.08 \rightarrow 1.14$ & $83\%$ & 29, 55, 85 & 0.78 & 1.00 \\
ESO~2.2\,m \#1  & 2012 04 02 & 03:34 & 06:56 & 61 & 120 & 200 & Sloan $i^{\prime}$ & $1.25 \rightarrow 1.08 \rightarrow 1.14$ & $83\%$ & 29, 55, 85 & 0.88 & 1.06 \\
ESO~2.2\,m \#1  & 2012 04 02 & 03:34 & 06:56 & 61 & 120 & 200 & Sloan $z^{\prime}$ & $1.25 \rightarrow 1.08 \rightarrow 1.14$ & $83\%$ & 30, 55, 85 & 0.99 & 1.15 \\[2pt]
ESO~2.2\,m \#2  & 2012 04 17 & 03:38 & 09:26 &138 & 124 & 150 & Sloan $g^{\prime}$ & $1.13 \rightarrow 1.08 \rightarrow 2.70$ & $ 9\%$ & 29, 55, 85 & 0.96 & 1.00 \\
ESO~2.2\,m \#2  & 2012 04 17 & 03:38 & 09:26 &138 & 124 & 150 & Sloan $r^{\prime}$ & $1.13 \rightarrow 1.08 \rightarrow 2.70$ & $ 9\%$ & 34, 55, 85 & 0.74 & 1.42 \\
ESO~2.2\,m \#2  & 2012 04 17 & 03:38 & 09:26 &138 & 124 & 150 & Sloan $i^{\prime}$ & $1.13 \rightarrow 1.08 \rightarrow 2.70$ & $ 9\%$ & 33, 55, 85 & 0.84 & 1.55 \\
ESO~2.2\,m \#2  & 2012 04 17 & 03:38 & 09:26 &138 & 124 & 150 & Sloan $z^{\prime}$ & $1.13 \rightarrow 1.08 \rightarrow 2.70$ & $ 9\%$ & 30, 55, 85 & 1.01 & 1.46 \\[2pt]
ESO~2.2\,m \#3  & 2012 04 21 & 04:28 & 08:58 &102 & 124 & 150 & Sloan $g^{\prime}$ & $1.08 \rightarrow 2.46$                  & $ 1\%$ & 28, 55, 85 & 3.26 & 1.60 \\
ESO~2.2\,m \#3  & 2012 04 21 & 04:28 & 08:58 &102 & 124 & 150 & Sloan $r^{\prime}$ & $1.08 \rightarrow 2.46$                  & $ 1\%$ & 35, 60, 90 & 0.77 & 1.40 \\
ESO~2.2\,m \#3  & 2012 04 21 & 04:28 & 08:58 &102 & 124 & 150 & Sloan $i^{\prime}$ & $1.08 \rightarrow 2.46$                  & $ 1\%$ & 32, 60, 90 & 0.93 & 1.32 \\
ESO~2.2\,m \#3  & 2012 04 21 & 04:28 & 08:58 &102 & 124 & 150 & Sloan $z^{\prime}$ & $1.08 \rightarrow 2.46$                  & $ 1\%$ & 32, 60, 90 & 1.02 & 1.12 \\[2pt]
CAHA~2.2\,m     & 2012 05 09 & 21:25 & 01:43 & 89 & 120 & 200 & Gunn $g$           & $1.50 \rightarrow 1.40 \rightarrow 2.15$ & $79\%$ & 15, 55, 80 & 1.65 & 1.18 \\
CAHA~2.2\,m     & 2012 05 09 & 21:25 & 01:43 & 89 & 120 & 200 & Gunn $r$           & $1.50 \rightarrow 1.40 \rightarrow 2.15$ & $79\%$ & 15, 50, 80 & 0.92 & 1.17 \\
CAHA~2.2\,m     & 2012 05 09 & 21:25 & 01:43 & 89 & 120 & 200 & Gunn $z$           & $1.50 \rightarrow 1.40 \rightarrow 2.15$ & $79\%$ & 20, 40, 55 & 1.70 & 1.01 \\[2pt]
Cassini~1.52\,m & 2012 05 09 & 21:01 & 00:28 & 64 & 170 & 180 & Gunn $r$           & $1.64 \rightarrow 1.57 \rightarrow 2.11$ & $80\%$ & 28, 55, 85 & 0.80 & 1.12 \\
CAHA~1.23\,m    & 2013 02 16 & 01:53 & 05:54 &128 & 140 & 180 & Cousin $I$         & $1.74 \rightarrow 1.40 \rightarrow 1.55$ & $33\%$ & 20, 45, 60 & 1.36 & 1.26 \\
\hline
\end{tabular}
\end{table*}

\subsection{Telescope defocussing observations of planetary transits}
\label{sec:2.1}

All the observations presented in this work were performed using
the \emph{telescope-defocussing} technique
\citep{alonso2008,southworth2009al1}. In this method the telescope
is defocussed so point spread functions (PSFs) cover of order 1000
pixels, and long exposure times (up to $\sim120$\,s) are used to
collect many photons in each PSF. This increases the
observational efficiency as the CCD is read out less often, thus
minimising Poisson and scintillation noise. The large PSFs are
also insensitive to focus or seeing changes, which might otherwise
cause systematic errors. The other main source of systematic
error, flat-fielding, is decreased by two orders of magnitude as
each PSF covers of order $10^3$ pixels. Telescope pointing
errors affect photometry via flat-fielding errors, so these also
average down to very low levels.

The exposure time is chosen for a given observing sequence from
consideration of the brightness of the target and comparison
stars, sky background, telescope size and filter used. The amount
of defocussing is then tuned so the peak count rate in the PSFs of
the target and comparison stars is significantly below the onset
of nonlinearity effects in the CCD. Changes in seeing, airmass and
sky transparency affect the count rate of the observations; this
is accounted for by changing the exposure times but not the focus
setting during an observing sequence.

\subsection{Two-site observations of planetary transits}
\label{sec:2.2}

Time-series photometry of transit events can show anomalies due to
the planet crossing over spots on the stellar surface. The
detection of starspots occulted by a transiting planet is becoming
commonplace
\citep[e.g.][]{pont2007,rabus2009,silva2010,desert2011,sanchis2011a,sanchis2011b,tregloan2013,mohler2013,mancini2013c}.
However, in the case of ground-based observations, similar signals
could be caused by weather-related or instrumental effects. One
method to sift the astrophysical from observational anomalies is
to observe a transit event from multiple telescopes at different
observatories. Any feature present in all light curves is
unambiguously intrinsic to the target of the observations. We used
this strategy to observe a transit of Qatar-2 using two telescopes
at different locations.

This two-site observational strategy was successfully tested in
the follow-up of HAT-P-8, where an anomaly was detected in both
the light curves \citep{mancini2013a}. It was also used for
HAT-P-16 and WASP-21 \citep{ciceri2013}, although in these two
cases no anomalies were detected. Conversely, \citet{lendl2013}
used this method to show that a possible starspot anomaly in the
WASP-19 system was of instrumental origin.

\subsection{Multi-band observations of planetary transits}
\label{sec:2.3}

Precise photometric observations of planetary transits probe the
chemical composition of the atmosphere of TEPs in a way similar to
transmission spectroscopy. A dependence of opacity on wavelength
causes variations in the radius of the planet as found from
transit observations. The effect can be big enough to measure
using medium-size telescopes with multi-band imagers, assuming
they have a good spectral resolution. It is important to obtain
the observations at multiple wavelengths simultaneously, to avoid
variations in transit depth due to unseen starspots rather than
planetary radius variations, even if one should take into account
that unocculted star spots may still cause wavelength dependence
of the transit depth \citep{sing2011}. In order to investigate
this effect, one should monitor the variability of the parent star
for many years\footnote{As an example, in the case of the K-dwarf
HD\,189733\,A, \citet{pont2013} estimated $0.3\%$ at $8\mu$m (and
then scaled at other wavelengths) as an additional uncertainty in
the depth measurement of individual transits due to unidentified
spot crossings.} or, assuming that stellar activity does not
change suddenly, repeatedly measure the transit depth by observing
several planetary-transit events a few days away from each other.

Simultaneous multi-band observations also allow a detailed study
of starspots which are occulted by the transiting planet. For a
single light curve the spot radius is strongly correlated with its
temperature \citep[e.g.][]{tregloan2013}. Multi-band light curves
constrain the spot temperature relative to the effective
temperature (\Teff) of the pristine photosphere, thus providing
additional information which lifts this degeneracy.

Simultaneous multi-band observations of planetary transits have
been obtained for several TEP systems using the instruments BUSCA
\citep{southworth2012al2,mancini2013a}, GROND
\citep{demooij2012,mancini2013b,mancini2013c,nikolov2013,southworth2013al,penev2013,mohler2013,bayliss2013},
ULTRACAM \citep{copperwheat2013,bento2013} and SIRIUS
\citep{narita2013}.

\begin{figure}
\centering
\includegraphics[width=9.cm]{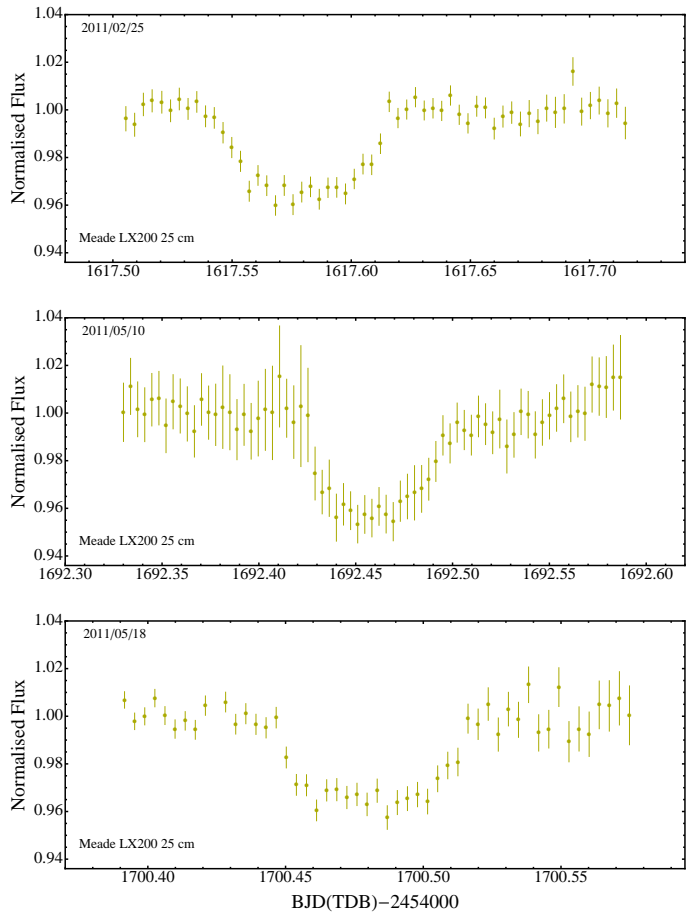}
\caption{Three transit events of Qatar-2\,b observed in 2011 with
a 25-cm MEADE LX200 telescope at the Canis Mayor observatory.}
\label{Fig:04}
\end{figure}

\section{Observations and data reduction}
\label{sec:3}

Five transits of Qatar-2\,b were monitored at optical wavelengths
by five different telescopes in 2012 and 2013
(Table\,\ref{Table:1}). One transit was followed simultaneously by
two of the telescopes, one was simultaneously observed through
three filters, and the other three through four filters. All
observations were performed with autoguiding and defocussing. The
light curves are given in Table\,\ref{Table:2} and shown in Figs.\
\ref{Fig:01}, \ref{Fig:02} and \ref{Fig:03}. We obtained
observations of three more transits in 2011 with a 25\,cm
telescope (Fig.\,\ref{Fig:04}).

All observations were analysed with the {\sc defot} pipeline
\citep{southworth2009al1} written in {\sc idl}\footnote{The
acronym {\sc idl} stands for Interactive Data Language and is a
trademark of ITT Visual Information Solutions.}. Debiasing and
flat-fielding were done using master calibration frames obtained
by median-combining individual calibration images. Pointing
variations were corrected by cross-correlating each image against
a reference frame. No de-correlation with PSF location was
necessary because, due to defocussing, the PSFs are much bigger
than the pixel sizes of the CCDs and image motion as the
telescopes tracked. Apertures were placed by hand on the target
and comparison stars, and their radii were chosen based on the
lowest scatter achieved when compared with a fitted model.

The {\sc aper} routine\footnote{{\sc aper} is part of the {\sc
astrolib} subroutine library distributed by NASA.} used to measure
the differential photometry commonly returns underestimated
errorbars. We therefore enlarged the errorbars for each light
curve to give a reduced $\chi^{2}$ of $\chi_{\nu}^{2}=1$ versus a
fitted model. We then further inflated the errorbars using the
$\beta$ approach (e.g.\ \citealt{gillon2006,winn2008,gibson2008})
to account for any correlated noise. We calculated $\beta$ values
for between two and ten data points for each light curve, and
adopted the largest $\beta$ value. They are reported in
Table\,\ref{Table:1}.

The twelve GROND light curves are plotted according to date
(Fig.\,\ref{Fig:01}) and filter (Fig.\,\ref{Fig:02}) in order to
highlight the starspot anomalies found in each transit.

\begin{table}
\centering \tiny %
\caption{Excerpts of the light curves of
Qatar-2: this table will be made available at the CDS. A portion
is shown here for guidance regarding its form and content.}
\label{Table:2} %
\begin{tabular}{llcrc}
\hline
Telescope    & Filter & BJD(TDB) & Diff.\ mag. & Uncertainty  \\
\hline
ESO~2.2\,m \#1  & $g^{\prime}$ & 2456022.654686 &  0.000290 & 0.001370 \\
ESO~2.2\,m \#1  & $g^{\prime}$ & 2456022.656996 & -0.001152 & 0.001314 \\[2pt]
ESO~2.2\,m \#2  & $r^{\prime}$ & 2456034.658998 &  0.001033 & 0.001019 \\
ESO~2.2\,m \#2  & $r^{\prime}$ & 2456034.660816 &  0.000272 & 0.001015 \\[2pt]
ESO~2.2\,m \#3  & $i^{\prime}$ & 2456038.693584 &  0.001068 & 0.001281 \\
ESO~2.2\,m \#3  & $i^{\prime}$ & 2456038.695359 &  0.000085 & 0.001276 \\[2pt]
CAHA~2.2\,m     & $z$          & 2456057.399202 &  0.001767 & 0.002883 \\
CAHA~2.2\,m     & $z$          & 2456057.401301 &  0.001192 & 0.002921 \\[2pt]
Cassini~1.52\,m & $r$          & 2456057.382333 &  0.000839 & 0.001281 \\
Cassini~1.52\,m & $r$          & 2456057.385550 &  0.000361 & 0.001184 \\[2pt]
CAHA~1.23\,m    & $I$          & 2456339.583029 & -0.000127 & 0.001653 \\
CAHA~1.23\,m    & $I$          & 2456339.586323 & -0.000021 & 0.001636 \\
\hline
\end{tabular}
\end{table}

\subsection{MPG/ESO 2.2-m telescope}
\label{sec:3.1}

Three transits of Qatar-2\,b were monitored with the GROND
instrument mounted on the MPG\footnote{Max Planck
Gesellschaft.}/ESO 2.2\,m telescope at the ESO observatory in La
Silla, Chile. The transit events were observed on 2012 April 2, 17
and 21. GROND is an imaging system capable of simultaneous
photometric observations in four fixed optical (similar to Sloan
$g^{\prime}$, $r^{\prime}$, $i^{\prime}$, $z^{\prime}$) and three
fixed NIR ($J,\, H,\, K$) passbands \citep{greiner2008}. Each of
the four optical channels is equipped with a back-illuminated
$2048 \times 2048$ E2V CCD, with a field-of-view (FOV) of
$5.4^{\prime} \times 5.4^{\prime}$ at a scale of
$0.158^{\prime\prime}/\rm{pixel}$. The three NIR channels use
$1024 \times 1024$ Rockwell HAWAII-1 arrays with a FOV of
$10^{\prime}\times 10^{\prime}$ at
$0.6^{\prime\prime}/\rm{pixel}$. Unfortunately, due to a lack of
good reference stars in the FOV, we were not able to obtain usable
light curves in the three NIR bands.

The precision of the optical data are in agreement with the
statistical-uncertainty study performed by \citet{pierini2012}
with two exceptions. The $z^{\prime}$ transit observed on 2012
April 2 has a lower depth compared to the other light curves of
the same transit (upper panel of Fig.\,\ref{Fig:01}) or the
$z^{\prime}$ light curves of the other two transits (bottom panel
of Fig.\,\ref{Fig:02}). This was caused by an unknown instrumental
error, which could not be reliably corrected for during the data
reduction. The $g^{\prime}$ data observed on 2012 April 21 were
affected by excess readout noise, caused by another unknown
instrumental problem, so are very inaccurate compared to the other
three light curves of this transit (bottom panel of
Fig.\,\ref{Fig:01}) or to the $g^{\prime}$ light curves of the
other two transits (upper panel of Fig.\,\ref{Fig:02}).

\subsection{CAHA 2.2-m telescope}
\label{sec:3.2}

We observed one full transit on the night of 2012 May 9, using the
CAHA 2.2-m telescope and BUSCA imager at the German-Spanish
Astronomical Center at Calar Alto in Spain. BUSCA is designed for
simultaneous four-colour photometry and, unlike GROND, the user
has a choice of filters available for each arm. Each of the four
optical channels is equipped with a Loral CCD4855 4k$\times$4k CCD
with 15$\times$15\,$\mu$m pixels, providing an astronomical FOV of
nearly 12$\times$12 arcmin.

For our observations we selected a Str\"{o}mgren $u$ filter in the
bluest arm and standard Calar Alto Gunn $g$, $r$ and $z$ filters
in the other three arms. This choice led to a reduced field of
view (from $12^{\prime} \times 12^{\prime}$ to a circle of
$6^{\prime}$ in diameter), but had two advantages. Firstly the $g$
and $r$ filters have a much better throughput compared to the
default Str\"{o}mgren $b$ and $y$ filters. Secondly, the different
filter thicknesses meant the $u$ band, where the target star is comparatively faint, was less defocussed. The
CCDs were binned $2 \times 2$ to shorten the readout time. The
autoguider was operated in focus. Unfortunately, the data obtained
in the $u$ band were too strongly affected by atmospheric
extinction and poor signal-to-noise ratio to be useful. The $g$,
$r$ and $z$ light curves are plotted in Fig.\,\ref{Fig:03}.

\subsection{Cassini 1.52-m telescope}
\label{sec:3.3}

The transit event of 2012 May 9 was also observed with the BFOSC
(Bologna Faint Object Spectrograph \& Camera) imager mounted on
the 1.52\,m Cassini Telescope at the Astronomical Observatory of
Bologna in Loiano, Italy. The transit was not fully covered due to
the pointing limits of the telescope. The CCD was used unbinned,
giving a plate scale of $0.58^{\prime\prime}/\rm{pixel}$, for a
total FOV of $13^{\prime} \times 12.6^{\prime}$. A Gunn $r$ filter
was used. The CCD was windowed to decrease the readout time and
the telescope was autoguided and defocussed, allowing low scatter to be obtained
even though the observations were conducted at high
airmass (see Table\,\ref{Table:1}). The light curve is plotted in
Fig.\,\ref{Fig:03}. The Cassini data are consistent with the
presence of the starspot anomaly in the final phase of the transit
ingress.

\subsection{CAHA 1.23-m telescope}
\label{sec:3.4}

Another complete transit event was observed with the CAHA 1.23\,m
telescope, on the night of 2013 February 16. Mounted in the Cassergrain focus of
this telescope is the DLR-MKIII camera, which has $4000\times4000$
pixels, a plate scale of 0.32 arcsec pixel$^{-1}$ and a large FOV
of $21.5^{\prime}\times21.5^{\prime}$. The transit was monitored
through a Cousins-$I$ filter, the telescope was autoguided and
defocussed, and the CCD was windowed. The resulting light curve is
plotted in Fig.\,\ref{Fig:03}. An anomaly is also visible in this
light curve, shortly after the transit midpoint.

\subsection{Canis Mayor Observatory}
\label{sec:3.5}

Three complete transits were observed at the Canis Mayor
Observatory, located in Castelnuovo Magra, Italy. The instrument
used for the observations was a Meade LX200 GPS 10\,inch
telescope, equipped with an f/6.3 focal reducer and an SBIG ST8
XME CCD camera. Science frames were taken through the Baader
Yellow 495 Longpass filter and the exposure time was 300\,s. The
telescope was autoguided and slightly defocussed. The  light
curves are plotted in Fig.\,\ref{Fig:04}.

\section{Light-curve analysis}
\label{sec:4}
All of our high-precision light curves show possible starspot
crossing events, which must be analysed using a self-consistent
and physically realistic model. We use the {\sc
prism}\footnote{Planetary Retrospective Integrated Star-spot
Model.} and {\sc gemc}\footnote{Genetic Evolution Markov Chain.}
codes \citep{tregloan2013} for this. We have previously used these
codes to model HATS-2 \citep{mohler2013} and WASP-19
\citep{mancini2013c}.

{\sc prism} models planetary transits with starspot crossings
using a pixellation approach in Cartesian coordinates. {\sc gemc}
uses a Differential Evolution Markov Chain Monte Carlo (DE-MCMC)
approach to locate the parameters of the {\sc prism} model which
best fit the data, using a global search. {\sc prism} uses the
fractional radii, $r_{\mathrm{A}} = \frac{R_{\mathrm{A}}}{a}$ and
$r_{\mathrm{b}} = \frac{R_{\mathrm{b}}}{a}$, where
$R_{\mathrm{A}}$ and $R_{\mathrm{b}}$ are the true radii of the
star and planet, and $a$ is the orbital semimajor axis.

The fitted parameters of {\sc prism} are the sum and ratio of the
fractional radii ($r_{\mathrm{A}}+r_{\mathrm{b}}$ and
$k=\frac{r_{\mathrm{b}}}{r_{\mathrm{A}}}$), the orbital period and
inclination ($P$ and $i$), the time of transit midpoint ($T_{0}$)
and the two coefficients of the quadratic limb darkening (LD) law
($u_{\mathrm{A}}$ and $v_{\mathrm{A}}$). Each starspot is
represented by the longitude and colatitude of its centre
($\theta$ and $\phi$), its angular radius ($r_{\mathrm{spot}}$)
and its contrast ($\rho_{\mathrm{spot}}$), the latter being the
ratio of the surface brightness of the starspot to that of the
surrounding photosphere.

The datasets obtained with the 25-cm telescope were modelled using
the much faster {\sc jktebop}\footnote{\textsc{jktebop} is written
in FORTRAN77 and is available at: {\tt
http://www.astro.keele.ac.uk/jkt/codes/jktebop.html}} code, as no
starspot anomalies are visible. The parameters used for {\sc
jktebop} were the same as for {\sc prism}.

\subsection{Orbital period determination}
\label{sec:4.1}

{\tiny
\begin{table}
\centering \small %
\caption{Times of transit midpoint of Qatar-2\,b and their
residuals. TRESCA refer to the ``TRansiting ExoplanetS and
CAndidates''. {\bf References:} (1) Canis Major Observatory (this
work); (2) \citet{bryan2012}; (3) Strajnic et al. (TRESCA); (4)
Zibar M. (TRESCA); (5) Gonzales J. (TRESCA); (6) MPG/ESO 2.2-m
$g^{\prime}$ (this work); (7) MPG/ESO 2.2-m $r^{\prime}$ (this
work); (8) MPG/ESO 2.2-m $i^{\prime}$ (this work); (9) MPG/ESO
2.2-m $z^{\prime}$ (this work); (10) Dax T. (TRESCA); (11) Masek
M. (TRESCA); (12) Carreno A. (TRESCA); (13) Montigiani N., Manucci
M. (TRESCA); (14) Cassini 1.52-m (this work); (15) CAHA 2.2-m $g$
(this work); (16) CAHA 2.2-m $r$ (this work); (17) CAHA 2.2-m $z$
(this work); (18) Campbell J. (TRESCA); (19) CAHA 1.23-m (this
work); (20) Ren\'{e} R. (TRESCA); (21) Ayiomamitis A. (TRESCA);
(22) Jacobsen J. (TRESCA); (23) Kehusmaa P., Harlingten C.
(TRESCA); (24) Shadic S. (TRESCA); (25) Colazo C. et al.
(TRESCA).}
\label{Table:3}
\begin{tabular}{lrrc}
\hline
Time of minimum    & Cycle & Residual & Reference  \\
BJD(TDB)$-2400000$ & no.   & (JD)     &            \\
\hline
$55617.58156 \pm 0.00082 $ &  -5 &  0.00005 & 1    \\
$55624.26679 \pm 0.00011 $ &   0 & -0.00031 & 2    \\
$55692.46109 \pm 0.00270 $ &  51 &  0.00105 & 1    \\
$55700.47915 \pm 0.00083 $ &  57 & -0.00358 & 1    \\
$55974.59334 \pm 0.00072 $ & 262 &  0.00173 & 3    \\
$55978.60480 \pm 0.00120 $ & 265 &  0.00184 & 4    \\
$55986.62711 \pm 0.00069 $ & 271 &  0.00145 & 5    \\
$56022.72850 \pm 0.00018 $ & 298 &  0.00070 & 6    \\
$56022.72800 \pm 0.00017 $ & 298 &  0.00020 & 7    \\
$56022.72815 \pm 0.00016 $ & 298 &  0.00035 & 8    \\
$56022.72810 \pm 0.00022 $ & 298 &  0.00030 & 9    \\
$56026.73842 \pm 0.00033 $ & 301 & -0.00073 & 10   \\
$56030.75160 \pm 0.00150 $ & 304 &  0.00110 & 11   \\
$56034.76157 \pm 0.00010 $ & 307 & -0.00028 & 6    \\
$56034.76196 \pm 0.00012 $ & 307 &  0.00011 & 7    \\
$56034.76198 \pm 0.00015 $ & 307 &  0.00013 & 8    \\
$56034.76249 \pm 0.00015 $ & 307 &  0.00064 & 9    \\
$56037.43785 \pm 0.00068 $ & 309 &  0.00176 & 12   \\
$56038.77329 \pm 0.00010 $ & 310 &  0.00009 & 7    \\
$56038.77345 \pm 0.00012 $ & 310 &  0.00025 & 8    \\
$56038.77312 \pm 0.00015 $ & 310 & -0.00008 & 9    \\
$56045.45890 \pm 0.00042 $ & 315 &  0.00012 & 13   \\
$56057.49242 \pm 0.00018 $ & 324 & -0.00041 & 14   \\
$56057.49246 \pm 0.00028 $ & 324 & -0.00037 & 15   \\
$56057.49274 \pm 0.00016 $ & 324 & -0.00009 & 16   \\
$56057.49282 \pm 0.00023 $ & 324 & -0.00012 & 17   \\
$56336.94926 \pm 0.00059 $ & 533 & -0.00091 & 18   \\
$56339.62434 \pm 0.00027 $ & 535 &  0.00007 & 19   \\
$56343.63460 \pm 0.00110 $ & 538 & -0.00116 & 20   \\
$56394.44559 \pm 0.00056 $ & 576 & -0.00059 & 21   \\
$56410.49050 \pm 0.00100 $ & 588 & -0.00108 & 22   \\
$56411.83020 \pm 0.00058 $ & 589 & -0.00150 & 23   \\
$56419.85020 \pm 0.00130 $ & 595 & -0.00120 & 24   \\
$56442.58191 \pm 0.00014 $ & 612 & -0.00047 & 25   \\
\hline
\end{tabular}
\end{table}
}

\begin{figure*}
\centering
\includegraphics[width=16.cm]{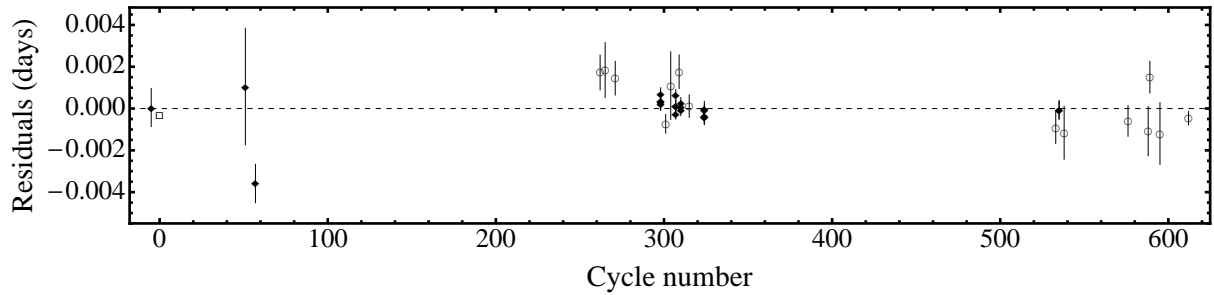}
\caption{Plot of the residuals of the times of transit midpoint of
Qatar-2\,b versus a linear ephemeris. The timings plotted with
diamonds are from this work, while empty circles are from ETD and
the box is from \citet{bryan2012}. The box has the same size of
the corresponding error bar, which has been suppressed for
clarity.} \label{Fig:05}
\end{figure*}

\begin{figure*}
\centering
\includegraphics[width=17.cm]{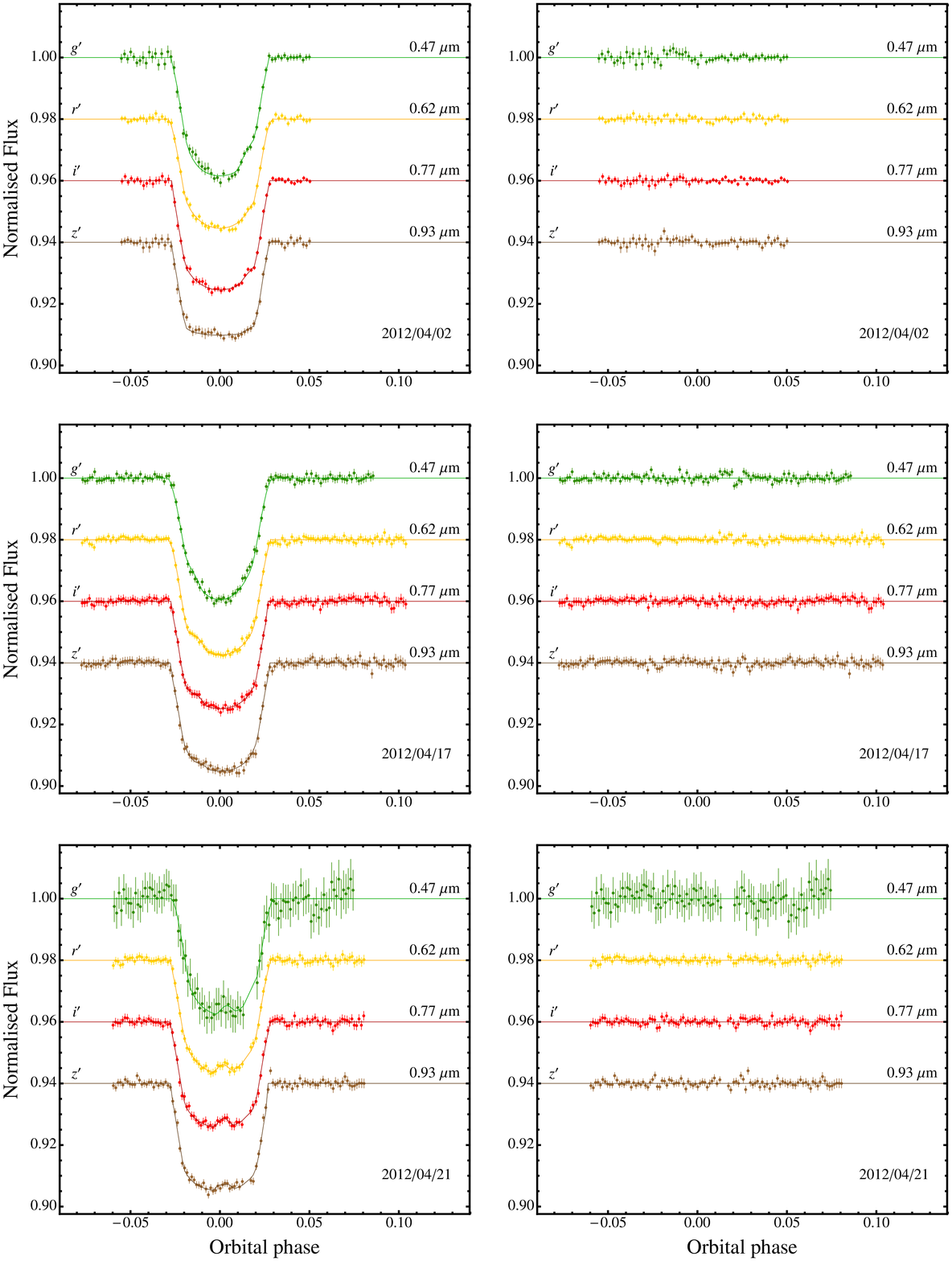}
\caption{Phased GROND light curves of Qatar-2 compared to the best
{\sc prism}+{\sc gemc} fits. The light curves and the residuals
are ordered according to the central wavelength of the filter
used. The passbands are labelled on the left of the figure, and
their central wavelengths are given on the right.} \label{Fig:06}
\end{figure*}

\begin{figure*}
\centering
\includegraphics[width=18.cm]{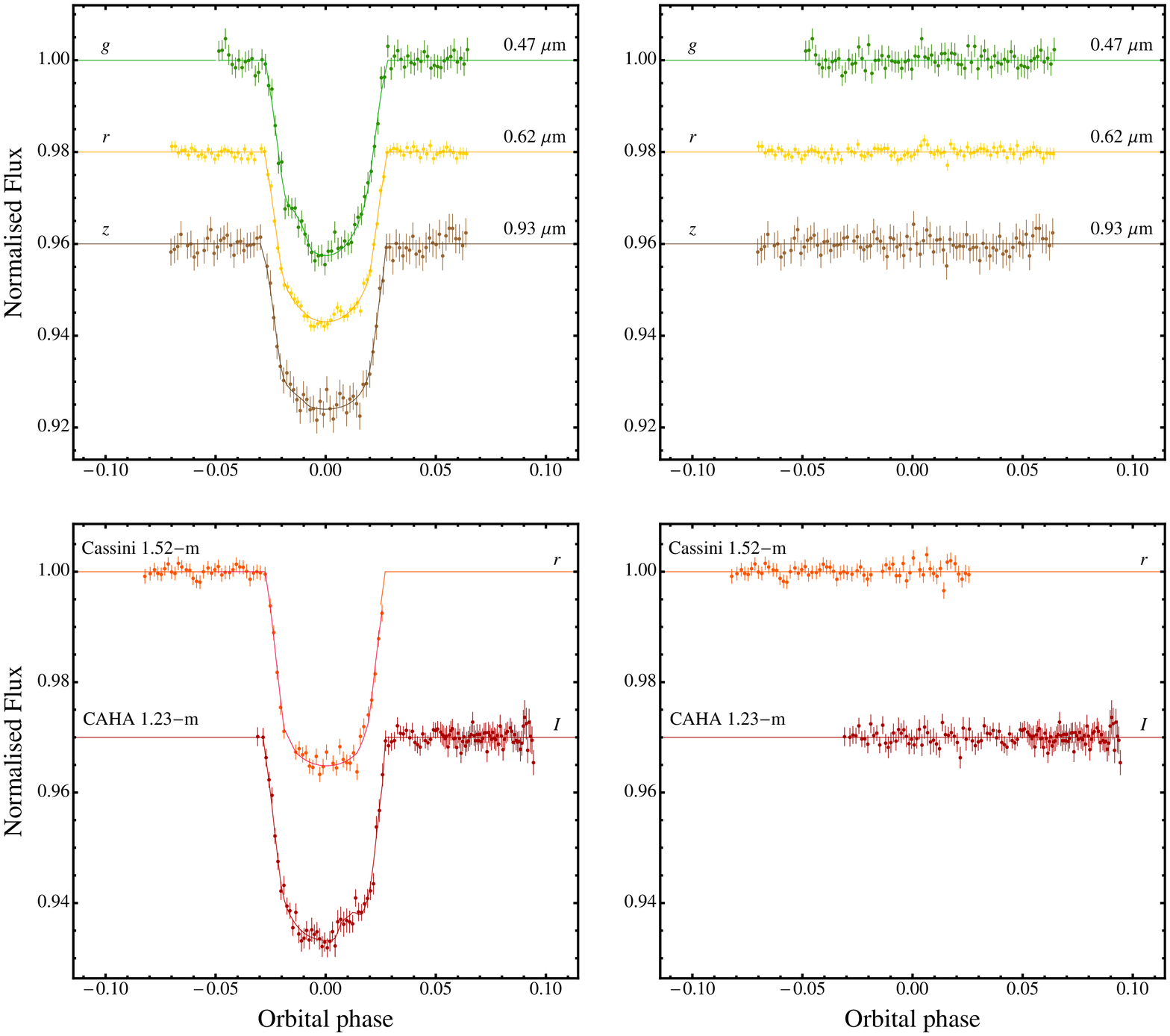}
\caption{\emph{Top panels}: phased BUSCA light curves of Qatar-2
compared to the best {\sc prism}+{\sc gemc} fits. The light curves
are ordered according to the central wavelength of the filter
used. The passbands are labelled on the left of the figure, and
their central wavelengths are given on the right. \emph{Bottom
panels}: as above, but for the Cassini 1.52-m and the CAHA 1.23-m
telescopes. The residuals of each fit are plotted in the right
panels.} \label{Fig:07}
\end{figure*}

We used the photometric data presented in Sect.\,\ref{sec:3} to
refine the orbital period of Qatar-2\,b. We excluded the third
$g^{\prime}$-band transit observed with GROND due to the large
scatter of these data. The transit times and uncertainties for the
high-precision datasets were obtained using {\sc prism}+{\sc
gemc}. Those for the small telescope were calculated using {\sc
jktebop} and Monte Carlo simulations. To these timings we added
one from the discovery paper \citep{bryan2012} and 14 measured by
amateur astronomers and available on the ETD\footnote{The
Exoplanet Transit Database (ETD) website can be found at
http://var2.astro.cz/ETD} website. The ETD light curves were
included only if they had complete coverage of the transit and a
Data Quality index $\leq 3$. All 34 timings were placed on the
BJD(TDB) time system (Table\,\ref{Table:3}).

Possible unknown planets in the system could gravitationally
perturb the orbit of Qatar-2\,b and induce transit timing
variations (TTVs). We therefore searched for periodic variations
in the transit times that might indicate such perturbations. We
first performed a weighted linear least-squares fit to compute a
new system ephemeris of $T_{\mathrm{T}} = T_0 + P \times E$, where
$E$ is the number of orbital cycles after the reference epoch and
\begin{equation}
T_{0} = \mathrm{BJD(TDB)} 2\,455\,624.267096 \pm 0.000087,
\nonumber
\end{equation}
\begin{equation}
P=1.33711647 \pm 0.00000026 \,\mathrm{d} \nonumber
\end{equation}
and the covariance between the two parameters is $-2.11 \times
10^{-11}$\,d$^2$. A plot of the residuals around the fit is shown
in Fig.\,\ref{Fig:05}. We then look for sinusoidal variations in
the residuals by scanning through a wide range of periods
(10--1000 orbits of Qatar-2\,b) and looking for the best-fitting
sinusoid at each period. Across this range of periods, the
best-fitting sinusoid has a period of 11.8\,d, but there are a
large range of local minima in the interval $15-200$ orbits. All
of these candidate TTV signals give $\chi^2 \sim 100$, implying
that our sinusoidal TTV models fit the data only poorly. The
semi-amplitudes of these best-fitting TTV models are all
$\sim$30\,sec, so we quote this value as the nominal upper limit
of any TTV effects on the orbit of Qatar-2b.

\subsection{Light-curve modelling}
\label{sec:4.2}

From this point we considered only the high-precision light curves
(i.e.\ not the Canis-Mayor ones). These were individually modelled
with {\sc prism}+{\sc gemc}, each time including the parameters
for one starspot. We used GEMC to randomly generate parameters for
36 chains, within a reasonable initial parameter space, and then
to simultaneously evolve the chains for 50\,000 successive
generations; see \citet{tregloan2013} for details. The light
curves and their best-fitting models are shown in Figs.\
\ref{Fig:06} and \ref{Fig:07}. The derived parameters of the
planetary system are reported in Table\,\ref{Table:4}, while those
of the starspots in Table\,\ref{Table:5}. We used the former to
reanalyse the phsyical properties of the system
(Sect.\,\ref{sec:5}), and the latter for the characterisation of
the starspots and the planetary orbit (Sect.\,\ref{sec:6}).

The results concerning the first GROND $z^{\prime}$ data set were
not considered since these data were compromised by an
instrumental error (Sect.\,\ref{sec:3.1}). Due to its low quality,
the starpsot parameters resulted from the fit of the third GROND
$g^{\prime}$ light curve have very large error bars and were not
reported in Table\,\ref{Table:5}.

We compared the fitted LD coefficients with the expected stellar
atmosphere model values. For the $g^{\prime}$, $r^{\prime}$,
$i^{\prime}$, $z^{\prime}$ and $I$ bands, we used the theoretical
LD coefficients estimated by \citet{claret2004} with two different
model atmosphere codes ({\sc atlas} and {\sc phoenix}). We also
checked the values for other passbands ($V$, $R_{\mathrm{c}}$,
$I_{\mathrm{c}}$, $R_{\mathrm{j}}$, $I_{\mathrm{j}}$), where
predictions from different authors
\citep{vanhamme1993,diaz1995,claret2000} are available. While
there is a good agreement for most of the light curves, there are
some, especially those related to the $g^{\prime}$ band, for which
differences of $\pm 0.1$ up to $\pm 0.3$ are apparent. Moreover,
one has also to consider that starspots do affect LD coefficients
as spots have different LD to the rest of the star
\citep{ballerini2012}. Based on these arguments, we could not
assume that theoretical LD coefficients are correct, and we
preferred to fit for the coefficients using {\sc prism}+{\sc
gemc}.

We discounted all the $g$-band data sets because the best-fitting
models returned very large value for $k$ compared to those of the
other redder bands. Considering the high activity level of
Qatar-2\,A, as can be seen from the numerous occulted starspots
detected, we attribute this to the effect of unocculted starspots
(i.e.\ starspots in a region of the stellar disc is not crossed by
the planet) which cause the transits monitored in the bluest bands
to be deeper. This is easily seen in Fig.\ 8 in which, using
Eq.\,(4) and (5) from \citet{sing2011}, we plot for the case of
Qatar-2 the correction for unocculted spots for a total dimming of
$1\%$ at a reference wavelength of 600\,nm for different starspot
temperatures. Starspots are modelled with {\sc atlas9} stellar
atmospheric models \citep{kurucz1979} of different temperatures
ranging from 4450 to 3700\,K in 250\,K intervals, and
$T_{\mathrm{eff}}=4645$\,K for the stellar temperature. However,
in order to apply the right correction for unocculted starspot, we
need an estimate of the absolute level of the stellar flux
corresponding to a spot-free surface, which can be obtained
through a continuing, accurate photometric monitoring of
Qatar-2\,A over several years. Since the QES discovery data are
not public and no other long photometric monitoring are available
for this target, we decided to exclude the $g$-band data for the
estimation of the physical parameters of the system. Instead,
corrections on the other optical bands ($\lambda>540$\,nm) are
expected to be of the order of $\lesssim 10^{-3}$, adding only a
small contribution to the uncertainties in the $k$ values.

\begin{figure*}%
\centering
\includegraphics[width=18.cm]{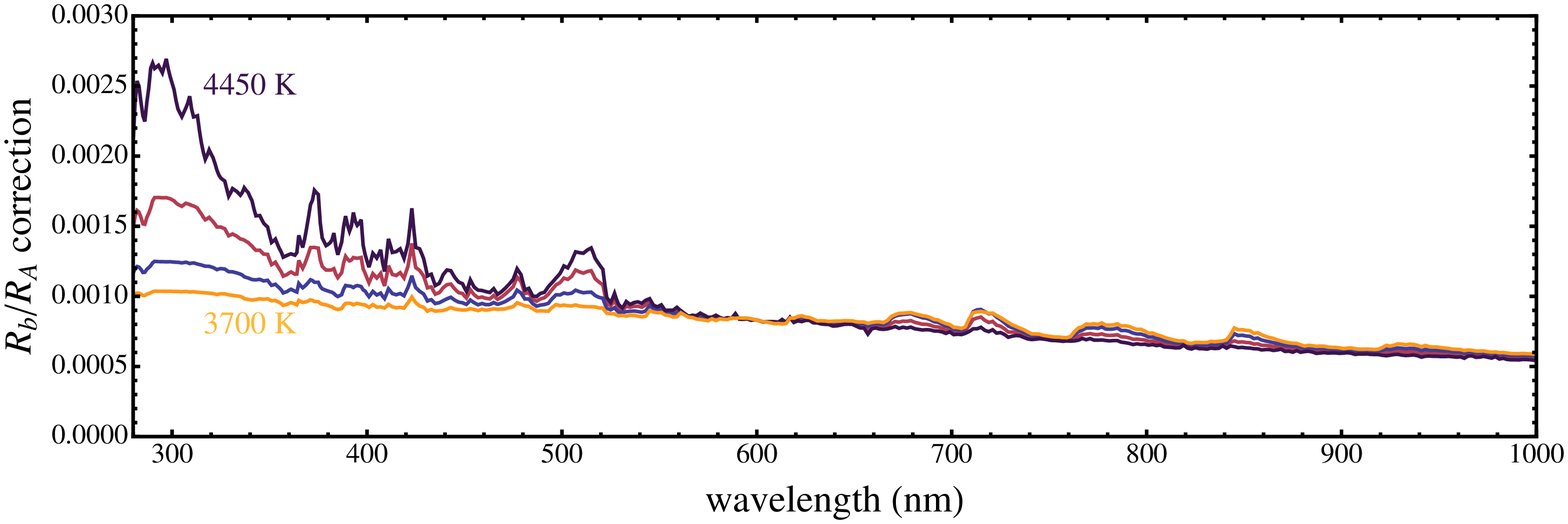}
\caption{The effect of unocculted starspots on the transmission
spectrum of Qatar-2 considering a $1\%$ flux drop at 600 nm. A
grid of {\sc atlas9} stellar atmospheric models at different
temperature, ranging from 4450 to 3700\,K in 250\,K intervals, was
used to model the starspot coverage, while for the star a model
with $T_{\mathrm{eff}}=4645$\,K was adopted.} \label{Fig:08}
\end{figure*}

\begin{table*}
\caption{Parameters of the {\sc prism}+{\sc gemc} best fits of the
light curves of Qatar-2 for the quadratic LD law with the
coefficients included as fitted parameters. The final parameters,
given in bold, are the weighted means of the results for the
individual datasets, but excluding the $g$-band ones (see text).
Results from the discovery paper are included at the base of the
table for comparison. The orbital period for each data set is
reported in Table\,\ref{Table:3}. The results concerning the GROND
$z^{\prime}$ dataset of the first transit are
not reported (see text).} %
\label{Table:4}
\centering
\begin{tabular}{llccccc}
\hline
Telescope & Filter & $r_{\mathrm{A}}+r_{\mathrm{b}}$ & $k$ & $i^{\circ}$ & $u_{\mathrm{A}}$ & $v_{\mathrm{A}}$ \\
\hline
ESO\,2.2-m \#1  & Sloan   $g^{\prime}$ & $0.1933 \pm 0.0045$ & $0.1695 \pm 0.0019 $ & $85.93 \pm 0.46$ & $0.275 \pm 0.039$ & $0.652 \pm 0.061$ \\
ESO\,2.2-m \#1  & Sloan   $r^{\prime}$ & $0.1964 \pm 0.0036$ & $0.1653 \pm 0.0016 $ & $85.62 \pm 0.39$ & $0.504 \pm 0.048$ & $0.187 \pm 0.022$ \\
ESO\,2.2-m \#1  & Sloan   $i^{\prime}$ & $0.1929 \pm 0.0022$ & $0.1663 \pm 0.0011 $ & $86.20 \pm 0.26$ & $0.431 \pm 0.044$ & $0.182 \pm 0.017$ \\[2pt]
%
ESO\,2.2-m \#2  & Sloan   $g^{\prime}$ & $0.2013 \pm 0.0070$ & $0.1720 \pm 0.0043 $ & $85.09 \pm 0.78$ & $0.712 \pm 0.106$ & $0.077 \pm 0.011$ \\
ESO\,2.2-m \#2  & Sloan   $r^{\prime}$ & $0.1935 \pm 0.0028$ & $0.1681 \pm 0.0018 $ & $86.05 \pm 0.35$ & $0.580 \pm 0.060$ & $0.111 \pm 0.015$ \\
ESO\,2.2-m \#2  & Sloan   $i^{\prime}$ & $0.1936 \pm 0.0025$ & $0.1642 \pm 0.0014 $ & $86.25 \pm 0.31$ & $0.349 \pm 0.044$ & $0.331 \pm 0.045$ \\
ESO\,2.2-m \#2  & Sloan   $z^{\prime}$ & $0.2003 \pm 0.0027$ & $0.1641 \pm 0.0016 $ & $86.00 \pm 0.28$ & $0.261 \pm 0.042$ & $0.564 \pm 0.074$ \\[2pt]
ESO\,2.2-m \#3  & Sloan   $g^{\prime}$ & $0.1956 \pm 0.0134$ & $0.1704 \pm 0.0057 $ & $85.42 \pm 1.23$ & $0.479 \pm 0.218$ & $0.249 \pm 0.229$ \\
ESO\,2.2-m \#3  & Sloan   $r^{\prime}$ & $0.1947 \pm 0.0020$ & $0.1658 \pm 0.0019 $ & $86.11 \pm 0.26$ & $0.588 \pm 0.073$ & $0.137 \pm 0.025$ \\
ESO\,2.2-m \#3  & Sloan   $i^{\prime}$ & $0.1960 \pm 0.0017$ & $0.1647 \pm 0.0017 $ & $85.89 \pm 0.18$ & $0.423 \pm 0.144$ & $0.141 \pm 0.235$ \\
ESO\,2.2-m \#3  & Sloan   $z^{\prime}$ & $0.1951 \pm 0.0015$ & $0.1670 \pm 0.0010 $ & $86.17 \pm 0.18$ & $0.291 \pm 0.038$ & $0.234 \pm 0.032$ \\[2pt]
CAHA\,2.2-m     & Gunn    $g$          & $0.1970 \pm 0.0056$ & $0.1733 \pm 0.0029 $ & $85.93 \pm 0.58$ & $0.629 \pm 0.060$ & $0.315 \pm 0.046$ \\
CAHA\,2.2-m     & Gunn    $r$          & $0.1948 \pm 0.0016$ & $0.1676 \pm 0.0012 $ & $86.17 \pm 0.18$ & $0.543 \pm 0.043$ & $0.144 \pm 0.023$ \\
CAHA\,2.2-m     & Gunn    $z$          & $0.1991 \pm 0.0053$ & $0.1671 \pm 0.0024 $ & $86.05 \pm 0.55$ & $0.260 \pm 0.055$ & $0.488 \pm 0.089$ \\[2pt]
Cassini\,1.52-m & Gunn    $r$          & $0.1926 \pm 0.0039$ & $0.1661 \pm 0.0023 $ & $86.15 \pm 0.43$ & $0.367 \pm 0.146$ & $0.250 \pm 0.243$ \\
CAHA\,1.23-m    & Cousins $I$          & $0.1992 \pm 0.0027$ & $0.1665 \pm 0.0025 $ & $86.19 \pm 0.35$ & $0.441 \pm 0.135$ & $0.337 \pm 0.200$ \\
\hline
Final results   & & $\mathbf{0.19549 \pm 0.00065}$ & $\mathbf{0.16590 \pm 0.00044}$ & $\mathbf{86.12 \pm 0.08}$ \\
 \hline
\citet{bryan2012} & & & $0.16508 \pm 0.00080$ & $88.30 \pm 0.94$ \\
 \hline
\end{tabular}
\end{table*}

\begin{table*}
\centering 
\caption{Starspot parameters derived from the {\sc prism}+{\sc
gemc} fitting of the transit light curves presented in this work.
\newline{$^{(a)}$The longitude of the centre of the spot is defined to be
$0^{\circ}$ at the centre of the stellar disc and can vary from
$-90^{\circ}$ to $90^{\circ}$. $^{(b)}$The colatitude of the
centre of the spot is defined to be $0^{\circ}$ at the north pole
and $180^{\circ}$ at the south pole. $^{(c)}$Angular radius of the
starspot; note that $90^{\circ}$ degrees covers half of stellar
surface. $^{(d)}$Spot contrast; note that 1.0 equals the
brightness of the surrounding photosphere. $^{(e)}$The temperature
of the starspots are obtained by considering the photosphere and
the starspots as black bodies (see text in Sect.\,\ref{sec:6}).
The results concerning the GROND $z^{\prime}$ dataset of the first
transit are not reported, because this dataset is affected by
correlated noise. Due to very large uncertainties, the results
concerning the GROND $g^{\prime}$ dataset of the third transit are
not reported. The results for the Cassini data set are also not
reported due to the large uncertainties of the parameters due to
the fact that the light-curve points are very scattered during the
transit time and the sampling is not so good.}}%
\label{Table:5}
\begin{tabular}{llccccc}
\hline
Telescope & Filter & $\theta (^{\circ})\,^{a}$ & $\phi(^{\circ})\,^{b}$ & $r_{\rm spot}(^{\circ})\,^{c}$ & $\rho_{\rm spot}\,^{d}$ & Temperature (K)$\,^{e}$   \\
\hline
ESO\,2.2-m \#1  & Sloan   $g^{\prime}$ & $37.67 \pm 2.65$ & $73.98 \pm 2.44 $ & $6.26 \pm 0.64$ & $0.392 \pm 0.066$ & $4068 \pm 99$ \\
ESO\,2.2-m \#1  & Sloan   $r^{\prime}$ & $36.50 \pm 6.12$ & $80.34 \pm 7.27 $ & $6.07 \pm 0.97$ & $0.540 \pm 0.091$ & $4137 \pm 130$ \\
ESO\,2.2-m \#1  & Sloan   $i^{\prime}$ & $37.71 \pm 2.53$ & $73.63 \pm 5.61 $ & $3.09 \pm 0.26$ & $0.579 \pm 0.052$ & $4096 \pm 89$ \\[2pt]
ESO\,2.2-m \#2  & Sloan   $g^{\prime}$ & $-32.70 \pm 3.38$ & $66.50 \pm 10.17 $ & $4.48 \pm 0.51$ & $0.468 \pm 0.049$ & $4166 \pm 72$ \\
ESO\,2.2-m \#2  & Sloan   $r^{\prime}$ & $-29.64 \pm 2.26$ & $70.88 \pm 4.89 $ & $4.23 \pm 0.48$ & $0.588 \pm 0.048$ & $4201 \pm 74$ \\
ESO\,2.2-m \#2  & Sloan   $i^{\prime}$ & $-32.62 \pm 3.70$ & $69.30 \pm 6.18 $ & $4.02 \pm 0.67$ & $0.626 \pm 0.087$ & $4167 \pm 134$ \\
ESO\,2.2-m \#2  & Sloan   $z^{\prime}$ & $-26.61 \pm 3.66$ & $67.78 \pm 7.35 $ & $3.24 \pm 0.43$ & $0.660 \pm 0.072$ & $4143 \pm 125$ \\[2pt]
%
ESO\,2.2-m \#3  & Sloan   $r^{\prime}$ & $ 4.15 \pm 0.48$ & $72.89 \pm 4.23 $ & $4.83 \pm 0.50$ & $0.566 \pm 0.091$ & $4172 \pm 127$ \\
ESO\,2.2-m \#3  & Sloan   $i^{\prime}$ & $ 4.70 \pm 2.39$ & $72.73 \pm 4.76 $ & $4.96 \pm 2.76$ & $0.556 \pm 0.183$ & $4061 \pm 290$ \\
ESO\,2.2-m \#3  & Sloan   $z^{\prime}$ & $ 4.54 \pm 0.62$ & $70.00 \pm 5.74 $ & $3.76 \pm 0.57$ & $0.653 \pm 0.116$ & $4132 \pm 196$ \\[2pt]
CAHA\,2.2-m     & Gunn    $g$          & $-30.48 \pm 4.13$ & $67.76 \pm 9.52 $ & $3.83 \pm 0.93$ & $0.422 \pm 0.102$ & $4078 \pm 145$ \\
CAHA\,2.2-m     & Gunn    $r$          & $-30.89 \pm 5.89$ & $72.86 \pm 9.73 $ & $3.45 \pm 1.63$ & $0.563 \pm 0.095$ & $4144 \pm 137$ \\
CAHA\,2.2-m     & Gunn    $z$          & $-29.59 \pm 6.83$ & $84.71 \pm 23.14 $ & $4.84 \pm 0.99$ & $0.702 \pm 0.184$ & $4217 \pm 292$ \\[2pt]
CAHA\,1.23-m    & Cousins $I$          & $25.29 \pm 9.22$ & $68.97 \pm 11.26 $ & $4.72 \pm 4.16$ & $0.601 \pm 0.216$ & $4120 \pm 332$ \\
\hline
\end{tabular}
\end{table*}

\section{Physical parameters of the planetary system}
\label{sec:5}

We measured the physical properties of the Qatar-2 system using
the \emph{Homogeneous Studies} approach (see
\citealt{southworth2012} and references therein). This methodology
makes use of the photometric parameters reported in
Table\,\ref{Table:4}, spectroscopic parameters from the discovery
paper (velocity amplitude $K_{\mathrm{A}}=558.7 \pm
5.9$\,m\,s$^{-1}$, effective temperature $T_{\mathrm{eff}}=4645
\pm 50$\,K, metallicity [Fe/H] $=-0.02 \pm 0.08$\,dex, and
eccentricity $e=0$; \citealp{bryan2012}) and theoretical stellar
models to estimate the properties of the system.

A value was estimated for $K_{\mathrm{b}}$, the velocity amplitude
of the planet, and the full system properties were calculated
using standard formulae. $K_{\mathrm{b}}$ was then iteratively
adjusted to find the best agreement between the observed $r_{\rm
A}$ and \Teff, and the values of $\frac{R_{\rm A}}{a}$ and \Teff\
predicted by a theoretical model of the calculated mass. This was
done for a grid of ages from zero to 5\,Gyr, and for five
different sets of theoretical models, specified in
Table\,\ref{Table:6}). We imposed an upper limit of 5\,Gyr because
the strong spot activity of the host star implies a young age. The
formal best fits are found at the largest possible ages (20\,Gyr
in this case), which implies that the spectroscopic properties of
the host star are not a good match for the stellar density
implicitly but strongly constrained by the transit duration
\citep{SeagerMallen03apj}. Given the large number of available
light curves, the discrepancy is best investigated by obtaining
new spectroscopic measurements of the atmospheric properties of
Qatar-2\,A.

We found a reasonably good agreement between the results for the
five different sets of theoretical stellar models (see
Table\,\ref{Table:6}). The Claret models are the most discrepant,
but not by enough to reject these results. We also used a
model-independent method to estimate the physical parameters of
the system, via a calibration based on detached eclipsing binary
stars of mass $<$3\Msun\ \citep{enoch2010,southworth2011}. These
empirical results match those found by using the stellar models.

The final set of physical properties was obtained by taking the
unweighted mean of the five sets of values obtained using stellar
models, and are reported in Table\,\ref{Table:7}. Systematic
errors were calculated as the standard deviation of the results
from the five models for each output parameter.
Table\,\ref{Table:7} also shows the physical properties found by
\citet{bryan2012}, which are of lower precision. Our measurement
of the stellar density is $\sim 3.7\sigma$ smaller than that found
by \citet{bryan2012}. The effect of this is to yield larger radius
measurements for both star and planet, and a 30\% lower planetary
density.

\begin{table*}
\flushleft \caption{Derived physical properties of the Qatar-2
planetary system.} \label{Table:6}
\begin{tabular}{l r@{\,$\pm$\,}l r@{\,$\pm$\,}l r@{\,$\pm$\,}l r@{\,$\pm$\,}l r@{\,$\pm$\,}l r@{\,$\pm$\,}l}
\hline
\ & \mc{This work} & \mc{This work} & \mc{This work} & \mc{This work} & \mc{This work} & \mc{This work} \\
\ & \mc{(dEB constraint)} & \mc{({\sf Claret} models)} & \mc{({\sf Y$^2$} models)} & \mc{({\sf Teramo} models)} & \mc{({\sf VRSS} models)} & \mc{({\sf DSEP} models)} \\
\hline
$K_{\rm b}$     (\kms) & 175.8    &   4.2     & 176.9    &   1.7     & 173.7    &   1.6     & 174.2    &   1.5     & 174.3    &   1.2     & 174.8    &   1.5     \\
$M_{\rm A}$    (\Msun) & 0.763    & 0.055     & 0.777    & 0.022     & 0.736    & 0.020     & 0.742    & 0.019     & 0.743    & 0.016     & 0.750    & 0.020     \\
$R_{\rm A}$    (\Rsun) & 0.783    & 0.019     & 0.788    & 0.008     & 0.774    & 0.008     & 0.775    & 0.007     & 0.776    & 0.006     & 0.778    & 0.007     \\
$\log g_{\rm A}$ (cgs) & 4.534    & 0.011     & 4.536    & 0.005     & 4.528    & 0.005     & 4.529    & 0.005     & 4.530    & 0.004     & 4.531    & 0.005     \\[2pt]
$M_{\rm b}$    (\Mjup) & 2.539    & 0.125     & 2.571    & 0.055     & 2.480    & 0.052     & 2.492    & 0.050     & 2.495    & 0.044     & 2.511    & 0.052     \\
$R_{\rm b}$    (\Rjup) & 1.265    & 0.031     & 1.273    & 0.013     & 1.250    & 0.012     & 1.253    & 0.012     & 1.254    & 0.010     & 1.258    & 0.012     \\
$\rho_{\rm b}$ (\pjup) & 1.173    & 0.034     & 1.166    & 0.022     & 1.187    & 0.022     & 1.184    & 0.022     & 1.183    & 0.021     & 1.180    & 0.022     \\
\safronov\             & 0.1142   & 0.0030    & 0.1135   & 0.0017    & 0.1156   & 0.0017    & 0.1153   & 0.0016    & 0.1152   & 0.0015    & 0.1149   & 0.0017    \\[2pt]
$a$               (AU) & 0.02172  & 0.00052   & 0.02186  & 0.00020   & 0.02147  & 0.00019   & 0.02152  & 0.00018   & 0.02154  & 0.00015   & 0.02160  & 0.00019   \\
Age              (Gyr) &       \mc{ }        & \erc{5.0}{0.0}{1.4} & \erc{5.0}{0.0}{0.0} & \erc{5.0}{0.0}{0.0} & \erc{5.0}{0.0}{0.0} & \erc{5.0}{0.0}{0.0} \\
\hline
\end{tabular}
\end{table*}

\begin{table*}
\centering %
\caption{Final physical properties of the Qatar-2 planetary
system, compared with results from \citet{bryan2012}. The first
error bar for each parameter is the statistical error, which stems
from the measured spectroscopic and photometric parameters. The
second error bar is the systematic error arising from the use of
theoretical stellar models, and is given only for those parameters
which have a dependence on stellar theory.} %
\label{Table:7}
\begin{tabular}{l l r@{\,$\pm$\,}c@{\,$\pm$\,}l r@{\,$\pm$\,}l}
\hline
\ & \ & \mcc{\bf This work (final)} & \mc{\citet{bryan2012}}  \\
\hline
Stellar mass                        & $M_{\rm A}$    (\Msun) & 0.743    & 0.020    & 0.007   & \mc{$0.740 \pm 0.037$}     \\
Stellar radius                      & $R_{\rm A}$    (\Rsun) & 0.776    & 0.007    & 0.003   & \mc{$0.713 \pm 0.018$}     \\
Stellar surface gravity             & $\log g_{\rm A}$ (cgs) & 4.530    & 0.005    & 0.001   & \mc{$4.601 \pm 0.018$}     \\
Stellar density                     & $\rho_{\rm A}$ (\psun) & \mcc{$1.591 \pm 0.016$}       & \mc{$ 2.05 \pm 0.12 $}     \\[2pt]
Planetary mass                      & $M_{\rm b}$    (\Mjup) & 2.494    & 0.052    & 0.016   & \mc{$2.487 \pm 0.086$}     \\
Planetary radius                    & $R_{\rm b}$    (\Rjup) & 1.254    & 0.012    & 0.004   & \mc{$1.144 \pm 0.035$}     \\
Planetary surface gravity           & $g_{\rm b}$     (\mss) & \mcc{$39.34 \pm  0.52$}       & \mc{$ 43.5 \pm 2.2  $}     \\
Planetary density                   & $\rho_{\rm b}$ (\pjup) & 1.183    & 0.022    & 0.004   & \mc{$ 1.66 \pm 0.13 $}     \\[2pt]
Planetary equilibrium temperature   & \Teq\              (K) & \mcc{$1344 \pm   14$}         & \mc{$ 1292 \pm 19   $}     \\
Safronov number                     & \safronov\             & 0.1152   & 0.0017   & 0.0004  & \mc{$0.740 \pm 0.037$}     \\
Orbital semimajor axis              & $a$               (AU) & 0.02153  & 0.00019  & 0.00007 & \mc{$0.02149 \pm 0.00036$} \\
\hline
\end{tabular}
\end{table*}

\section{Starspot modelling}
\label{sec:6}

As explained in Sect.\,\ref{sec:4}, the parameters of the
starspots detected in the light curves were fitted together with
those of the transit using {\sc prism} and {\sc gemc} codes. In
this way, we were able to establish the best-fitting position,
size, spot contrast and temperature for each of them. The results
are summarised in Table\,\ref{Table:5}. The final values for the
angular radii of the spots detected in each transit come from the
weighted mean of the results in each band. These are reported in
Table\,\ref{Table:8} in km and in percent of the stellar disc.

\begin{table*}
\centering%
\caption{Starspot sizes and temperatures for each of
the transits presented in this work.} \label{Table:8}
\begin{tabular}{lccc}
\hline
Telescope & Spot radius (km) & \% of the stellar disc & Temperature (K)   \\
\hline
ESO\,2.2-m \#1  & $34722 \pm 2224$ & $\sim0.41\%$ & $4094 \pm 59 $ \\
ESO\,2.2-m \#2  & $36951 \pm 2392$ & $\sim0.47\%$ & $4176 \pm 45 $ \\
ESO\,2.2-m \#3  & $41216 \pm 3528$ & $\sim0.58\%$ & $4148 \pm 100 $ \\
CAHA\,2.2-m     & $39352 \pm 5906$ & $\sim0.53\%$ & $4124 \pm 94 $ \\
CAHA\,1.23-m    & $70271 \pm 9628$ & $\sim1.70\%$ &$4120 \pm 332 $ \\
\hline
\end{tabular}
\end{table*}

Current knowledge on starspot temperatures is based on results
coming from different techniques, such as simultaneous modelling
of brightness and colour variations, Doppler imaging, modelling of
molecular bands and atomic line-depth ratios. Planetary-transit
events offer a more direct way to investigate this topic,
especially in the lucky case that the parent star is active and
that the planet occults one or more starspots during the transit.

For the current case, we observed starspots in every one of the
transits that we monitored with high precision. This suggests that
Qatar-2\,A could be in a peak of its stellar activity, since no
starspots were seen in the four light curves observed between
February and March 2011 by \citet{bryan2012}, of which three
covered the complete transit event.

\begin{figure*}%
\centering
\includegraphics[width=18.cm]{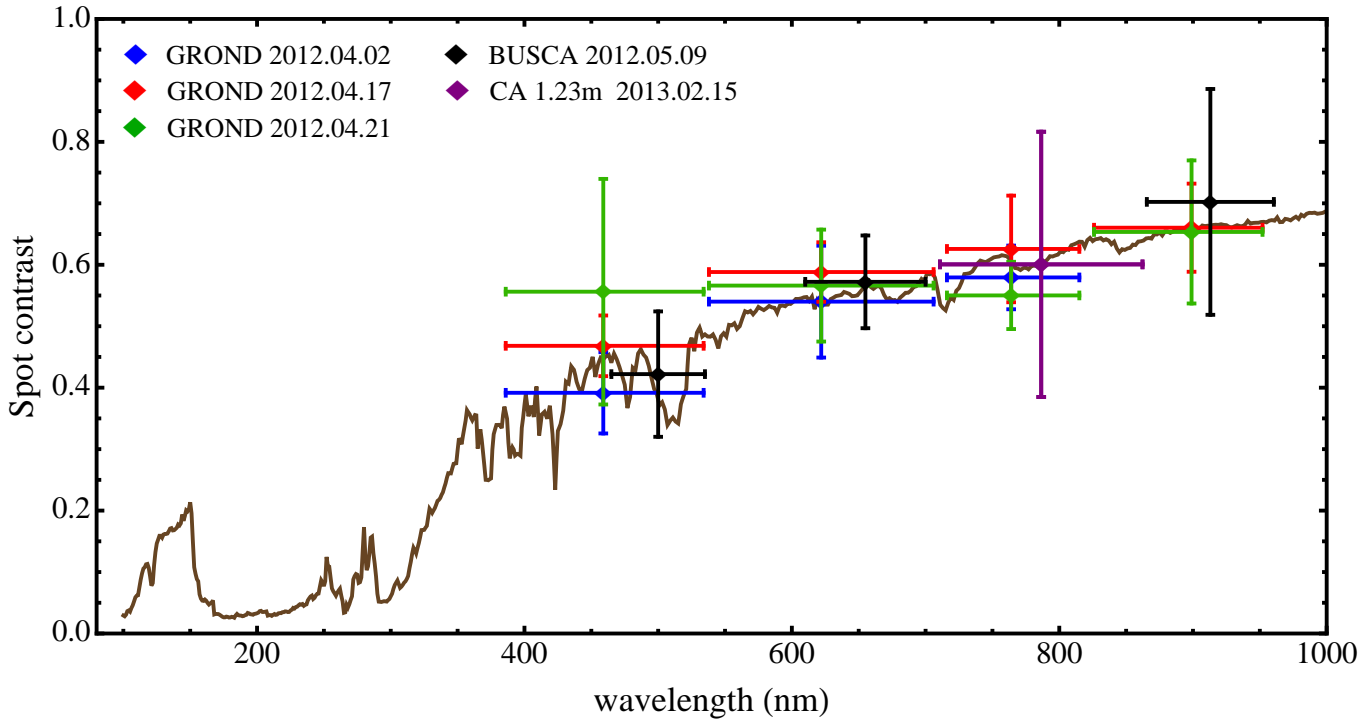}
\caption{Variation of the spot contrast with wavelength. All the
points are from this work and are explained in the upper-left
legend. The vertical bars represent the errors in the measurements
and the horizontal bars show the FWHM transmission of the
passbands used. Solid line represents the spot-contrast variation
expected for a starspot at 4200\,K over a stellar photosphere of
4645\,K.} \label{Fig:09}
\end{figure*}

\begin{figure*}%
\centering
\includegraphics[width=18.cm]{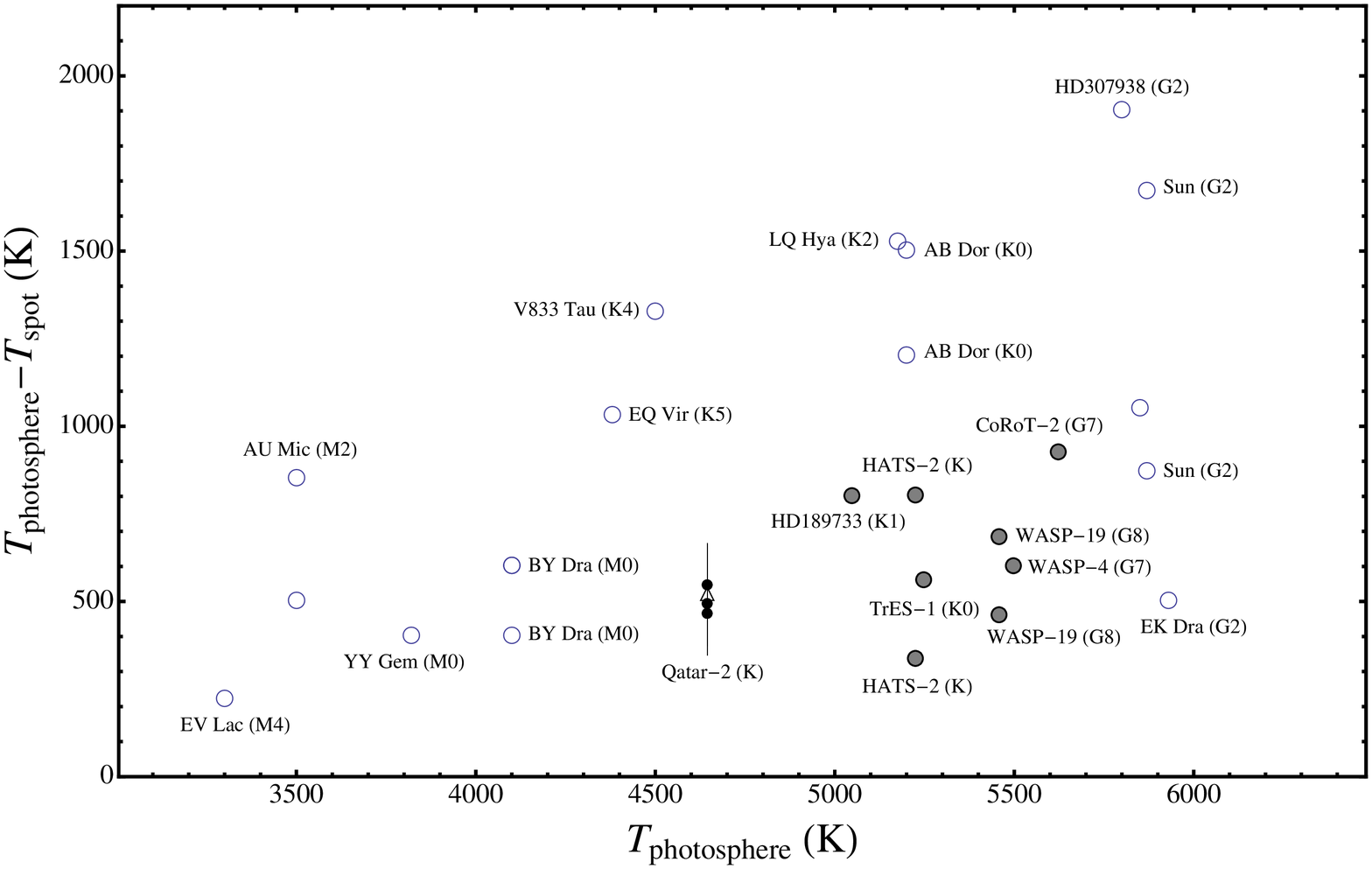}
\caption{Spot temperature contrast with respect to the
photospheric temperature in several dwarf stars taken from
\citet{berdyugina2005}, shown with empty circles. The name and
spectral type of the star are also reported for most of them. The
values for TrES-1, CoRoT-2, HD\,189733, WASP-4, HATS-2, WASP-19
are taken from \citet{rabus2009}, \citet{silva2010},
\citet{sing2011}, \citet{sanchis2011a}, \citet{mohler2013},
\citet{mancini2013c} and \citet{huitson2013}, respectively. Their
positions in the diagram are marked with gray dots. The error bars
have been suppressed for clarity. Note that some stars appear
twice. The black dots and the triangle refer to the values
estimated with GROND and BUSCA, respectively, for the case of
Qatar-2 (this work).} \label{Fig:10}
\end{figure*}

Taking advantage of our multi-band photometry, we studied how the
starspot contrast changes with passband. Starspots are expected to
be darker in the ultraviolet (UV) than in the infrared (IR). From
Table\,\ref{Table:5}, it is clear that the starspots are brighter
in the redder passbands than in the bluer passbands, for all four
simultaneous multi-band observations. Modelling both the
photosphere and the starspot as black bodies
\citep{rabus2009,sanchis2011a,mohler2013,mancini2013c} and using
Eq.\,1 of \citet{silva2003} and $T_{\mathrm{eff}}=4645\pm50$
\citep{bryan2012}, we estimated the temperature of the starspots
at different bands and reported them in the last column of Table
\ref{Table:5}. The values of the temperature estimated for each
transit are in good agreement between each other within the
experimental uncertainties and point to starspots with temperature
between 4100 and 4200 K. This can be also noted in
Fig.\,\ref{Fig:09}, where we compare the spot contrasts calculated
by {\sc prism+gemc} with those expected for a starspot at 4200\,K
over a stellar photosphere of 4645\,K, both modelled with {\sc
atlas9} atmospheric models \citep{kurucz1979}. The weighted means
are shown in Table \ref{Table:8}, and are consistent with what has
been observed for other main-sequence stars
\citep{berdyugina2005}, and for the case of the TrES-1
\citep{rabus2009}, CoRoT-2 \citep{silva2010}, HD\,189733
\citep{sing2011}, WASP-4 \citep{sanchis2011a}, HATS-2
\citep{mohler2013} and WASP-19 \citep{mancini2013c,huitson2013}.
All these measurements are shown in Fig.\,\ref{Fig:10} versus the
temperature of the photosphere of the corresponding star. The
spectral class for most of the stars is also reported and allows
to see that the temperature difference between photosphere and
starspots is not strongly dependent on spectral type, as already
noted by \citet{strassmeier2009}.

Another type of precious information that we can obtain from
follow-up transit observations comes from observing multiple
planetary transits across the same starspot or starspot complex
\citep{sanchis2011a}. In these cases, there is a good alignment
between the stellar spin axis and planet's orbital plane and, by
measuring the shift in position of the starspot between the
transit events, one can constrain the alignment between the
orbital axis of the planet and the spin axis of the star with
higher precision than with the measurement of the
Rossiter-McLaughlin effect (e.g.\ \citealt{tregloan2013}). On the
other hand, if the starspots detected at each transit are
different, then the latitude difference of the starspots is fully
degenerate with the sky-projected spin-orbit angle $\lambda$.

According to the orbital period of the transiting planet, the same
starspot can be observed after consecutive transits or after some
orbital cycles, presuming that in the latter case the star
performs one or more complete revolutions. Therefore, in general,
the distance $D$ covered by the starspot in the time between two
detections is
\begin{equation}
D=(n \times 2 \pi R_{\mathrm{lat}}) + d, %
\label{Eq:1}
\end{equation}
where $n$ is the number of revolutions completed by the star,
$R_{\mathrm{lat}}$ is the scaled stellar radius for the latitude
at which the starspot have been observed and $d$ is the arc length
between the positions of two transits in which the starspot is
detected.

In the present case, we observed three close transits of
Qatar-2\,b, with GROND (Fig.\,\ref{Fig:06}). The transits \#1 and
\#2 were separated by twelve days (nine cycles), while transits
\#2 and \#3 by four days (three cycles). Calculating the weighted
means of the starspot positions among the four bands for each of
the three transits, we estimate
\begin{eqnarray}
(\theta,\phi)_{\#1}&=& (37.59^{\circ} \pm 1.75^{\circ}, 74.48^{\circ} \pm 2.14^{\circ}) \nonumber \\
(\theta,\phi)_{\#2}&=&(-30.24^{\circ} \pm 1.52^{\circ}, 69.41^{\circ} \pm 3.22^{\circ}) \nonumber \\
(\theta,\phi)_{\#3}&=&  (4.22^{\circ} \pm 0.05^{\circ}, 72.16^{\circ} \pm 2.76^{\circ}) \nonumber \\
\end{eqnarray}
Could the starspots detected in the very close transits \#2 and
\#3 be the same? We find that Qatar-2\,A rotates unrealistically
slowly for the case $n=0$, i.e.\
$v_{(72^{\circ})}=0.31$\,km\,s$^{-1}$. For $n=1$ we get
$v_{(72^{\circ})}=9.6$\,km\,s$^{-1}$, accomplishing a complete
revolution in $P_{\mathrm{rot}}=3.88 \pm 0.07\,\mathrm{d}$ at a
colatitude of $72^{\circ}$. For $n=2$ we have
$v_{(72^{\circ})}=18.9$\,km\,s$^{-1}$, and for $n>2$ the star
would rotate even faster. The rotation period of Qatar-2\,A at the
equator, estimated from the sky-projected rotation rate and the
stellar radius \citep{sanchis2011a}, is
\begin{equation}
P_{\mathrm{rot}}\approx \frac{2 \pi
R_{\mathrm{A}}}{v\sin{i_{\star}}}= (14.0 \pm
2.5\,\mathrm{d})\sin{i_{\star}}, %
\label{Eq:2}
\end{equation}
where $i_{\star}$ is the inclination of the stellar rotation axis
with respect to the line of sight
($v\,\sin{i_{\star}}=2.8$\,km\,s$^{-1}$; \citealp{bryan2012}).
From this we can exclude that the starspots detected in transits
\#2 and \#3 are the same.


We now turn to the more promising transits \#1 and \#2.
From Eq.\,\ref{Eq:1} with $n=1$ we find $v_{(72^{\circ})}=3.28 \pm
0.13$\,km\,s$^{-1}$ which corresponds to $P_{\mathrm{rot}}=11.4
\pm 0.5\,\mathrm{d}$. This is within $1\sigma$ of the equatorial
value found above. Under the assumption that we have detected the
same spot in these two transits, simple algebra gives the
sky-projected angle between the stellar rotation and the planetary
orbit to be $\lambda=4.3^{\circ} \pm 4.5^{\circ}$. This is the
first measurement of the orbital obliquity of Qatar-2 and is
consistent with orbital alignment ($\lambda = 0$). This result is
also in agreement with the general idea that cool stars have low
obliquity \citep{winn2010}.

The fact that we observed spot crossing events in every one of our
high-precision transit light curves suggests that the host star
has an active region underneath the transit cord. This idea was
put forward by \citep{sanchis2011b} for HAT-P-11, a rather
different case where the planet's orbital axis is inclined by
nearly 90$^\circ$ to the stellar rotational axis and the spot
events cluster at two orbital phases in the transit light curve
\citep[see fig.\,24 in][]{southworth2011}.


\section{Variation of the planetary radius with wavelength}
\label{sec:7}

As discussed in Sect.\,\ref{sec:2.3}, simultaneous multi-band
transit observations allow the chemical composition of the
planet's atmosphere to be probed in a way similar to transmission
spectroscopy.

\begin{figure}
\centering
\includegraphics[width=9.cm]{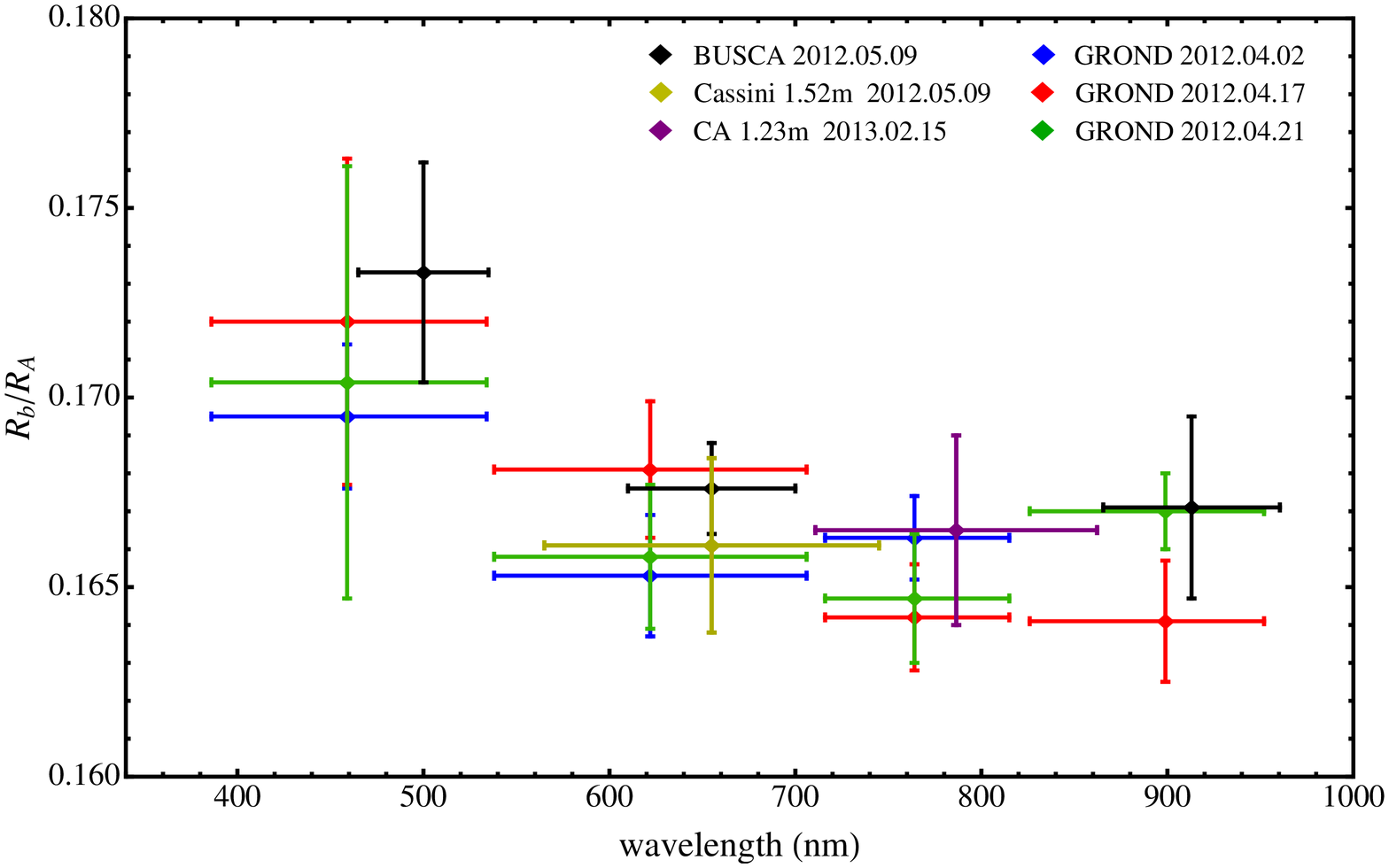}
\caption{Variation of the planetary radius, in terms of the
planet/star radius ratio, with wavelength. All the points are from
this work and are explained in the upper-right legend. The
vertical bars represent the errors in the measurements and the
horizontal bars show the FWHM transmission of the passbands used.}
\label{Fig:11}
\end{figure}

In Sect.\,\ref{sec:5} we estimated an equilibrium temperature of
$1344 \pm 14$\,K for Qatar-2\,b, which suggests, in the
terminology of \citet{fortney2008}, that this planet should belong
to the pL class. Therefore, based on the theoretical perspective
of \citet{fortney2008}, it is not expected that the atmosphere of
the planet should host a large amount of absorbing molecules, such
as gaseous titanium oxide (TiO) and vanadium oxide (VO). By using
the data reported in Table\,\ref{Table:4}, we investigated the
variations of the radius of Qatar-2\,b in the wavelength ranges
accessible to the instruments used, i.e.\ $386$--$976$\,nm. In
particular, we show in Fig.\,\ref{Fig:11} the values of $k$ (the
planet/star radius ratio) determined from the analysis of each
transit separately versus wavelength. The vertical bars represent
the relative errors in the measurements and the horizontal bars
show the full-width at half-maximum (FWHM) transmission of the
passbands used.

The depths from the transit observations all agree with each other
within $\sim$2$\sigma$, even if the $g$-band data indicate a
larger value of $k$ at this wavelength region. As discussed in
Sect.\,\ref{sec:4}, this is likely caused by unocculted starspots
which affect the measure of the transit depth in the bluest
optical bands (see Fig.\,\ref{Fig:08}).

\begin{figure}
\centering
\includegraphics[width=9.cm]{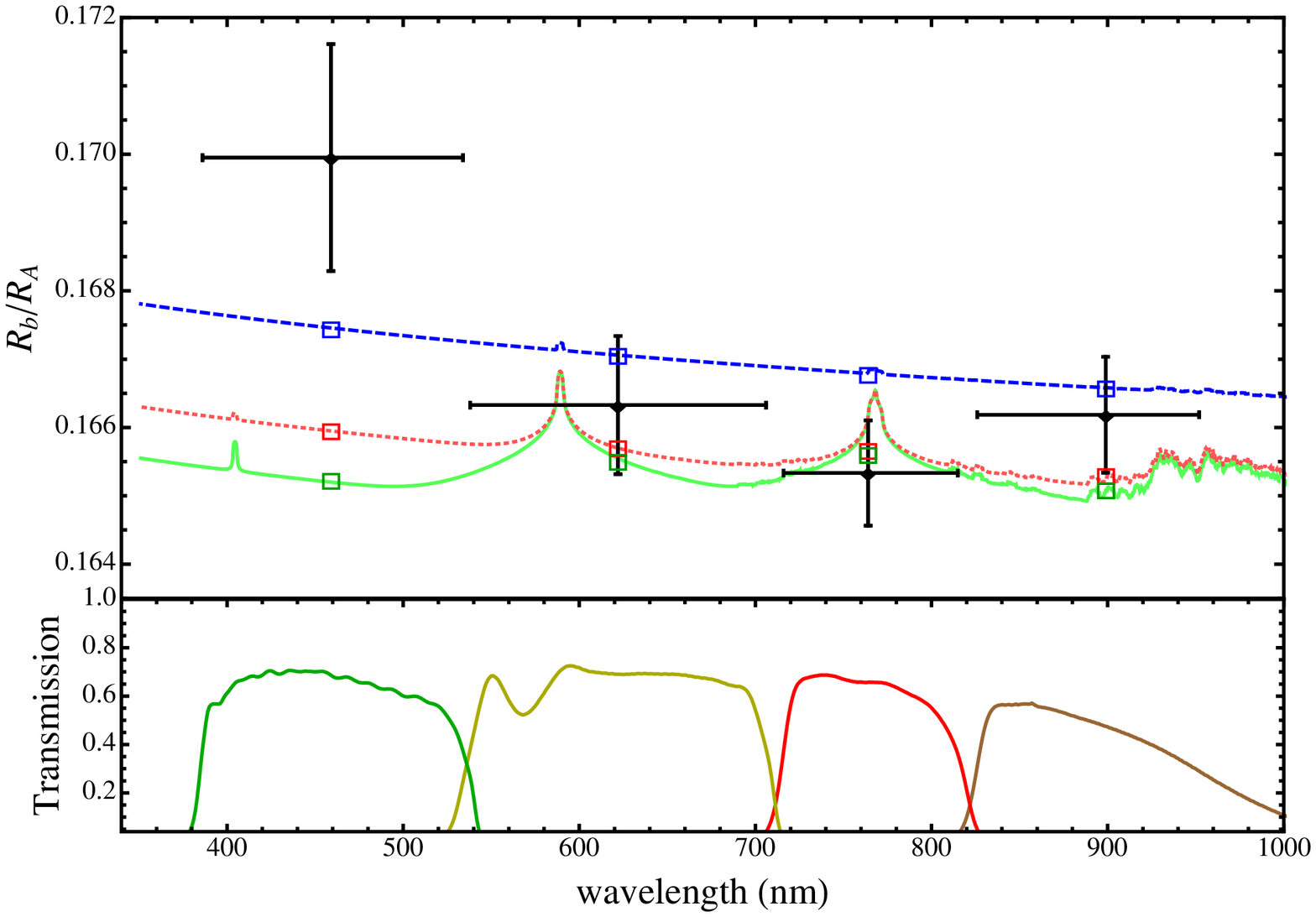}
\caption{Variation of the planetary radius, in terms of
planet/star radius ratio, with wavelength. The black points are
the weighted-mean results coming from the three transit
observations performed with GROND. The vertical bars represent the
errors in the measurements and the horizontal bars show the FWHM
transmission of the passbands used. The observational points are
compared with three synthetic spectra at 1250 K, which do not
include TiO and VO opacity. With respect to the model identified
with the green line, the other two have H$_{2}$/He Rayleigh
scattering increased by a factor of 10 (red dot line) and 1000
(blue dashed line). An offset is applied to all three models to
provide the best fit to our radius measurements. Coloured squares
represent band-averaged model radii over the bandpasses of the
observations. Transmission curves of the GROND filters are shown
in the bottom panel.} \label{Fig:12}
\end{figure}

In order to have a more homogeneous set of data, we consider only
the weighted-mean results coming from the three transit
observations performed with GROND over 19\,days. They are shown in
Fig.\,\ref{Fig:12} together with three one-dimensional model
atmospheres developed by \citet{fortney2010} for comparison. The
green line has been calculated for Jupiter's gravity (25\mss) with
a base radius of 1.25\Rjup\ at the 10\,bar level and at 1250\,K.
The opacity of TiO and VO molecules is excluded from the model and
the optical transmission spectrum is dominated by Rayleigh
scattering in the blue, and pressure-broadened neutral atomic
lines of sodium (Na) at 589\,nm and potassium (K) at 770\,nm. The
other two models are equal but with H$_{2}$/He Rayleigh scattering
increased by a factor of 10 (red dot line) and 1000 (blue dashed
line). The GROND data clearly indicate a very large radius of
Qatar-2\,b in the wavelength range $400$--$510$\,nm; the value of
$R_{\mathrm{b}}$ measured in the $g^{\prime}$ band differs by
$\sim11670$\,km from that in the $r^{\prime}$ band, which equates
to $\sim 60\,H$, where $H$ is the atmospheric pressure scale
height. A reasonable explanation for this nonphysical result is to
advocate the presence of unocculted starspots which strongly
contaminate the stellar flux in the $g^{\prime}$ band. Actually,
if we correct the value of $k$ in the $g^{\prime}$ band by 0.0015
(cfr. Fig.\,\ref{Fig:08}), we obtain a planetary radius which is
still large compared to the other bands, but straightforwardly
explicable within the error bar by Rayleigh-scattering processes
occurring in the atmosphere of Qatar-2\,b (blue line in
Fig.\,\ref{Fig:12}). A long photometric monitoring of Qatar-2\,A
is mandatory to estimate the right correction to make and new
transit events observations in the $u$ or $U$ bands are also
suggested in order to confirm the Rayleigh-scattering signature.

The lower value of the radius in the $i^{\prime}$ band could
suggest a lack of K, even if the spectral resolution of GROND is
not enough good to allow a clear determination. However, if true,
this lack is attributable to a selective destruction process via
photoionization, or its presence in a more condensed state.
Another possible explanation is that the planet formed very far
from the star, on the boundaries of the protoplanetary disk,
before migrating to its current position.

\section{Summary and conclusions}
\label{sec:8}

We have reported new broad-band photometric observations of five
transit events in the Qatar-2 planetary system. Three of them were
simultaneously monitored through four optical bands with GROND at
the MPG/ESO 2.2-m telescope, and one through three optical bands
with BUSCA at the CAHA 2.2-m telescope and contemporaneously with
the Cassini 1.52-m telescope in one band. Another single-band
observation was obtained at the CAHA 1.23-m telescope. In total we
have collected 17 new light curves, all showing anomalies due to
the occultation of starspots by the planet. These were fitted
using the {\sc prism}+{\sc gemc} codes which are designed to model
transits with starspot anomalies. Three further transits of
Qatar-2\,b were observed with a 25-cm telescope, and fitted with
the {\sc jktebop} code. Our principal results are as follows.

($i$) We used our new data and those collected from the discovery
paper and web archives to improve the precision of the measured
orbital ephemerides. We also investigated possible transit timing
variations generated by a putative outer planet and affecting
Qatar-2\,b's orbit. Unfortunately, the sampling of transit timings
is not yet sufficient to detect any clear sinusoidal signal. A
more prolonged monitoring of this system is mandatory in order to
accurately characterise a possible TTV signal.

($ii$) We have revised the physical parameters of Qatar-2\,b and
its host star, significantly improving their accuracy. They are
summarised in Table\,\ref{Table:7}. In particular we find that
density of Qatar-2\,b is lower than the estimate by
\citet{bryan2012}. The theoretical radius calculated by
\citet{fortney2007} for a core-free planet at age 4.5\,Gyr and
distance 0.02\,AU is 1.17\Rjup\ for a planet of mass 2.44\Mjup.
These numbers are in good agreement with the parameters that we
have estimated and imply that Qatar-2\,b is coreless.
Fig.\,\ref{Fig:11} shows the new position of Qatar-2\,b in the
mass-radius diagram (left-hand panel) and in the plot of orbital
period versus surface gravity (right-hand panel).

($iii$) The detection of so many starspots in our light curves
suggests an intense period of activity for Qatar-2\,A. The extent
of each starspot was estimated and found to be in agreement with
those found for similar starspot detections in other planetary
systems during transit events. The projected positions
($\theta,\phi$) of each starspot was also determined, and the
colatitudes were found to be consistent with $72^{\circ}$. For the
four simultaneous multi-band observations, we detected a variation
of the starspot contrast as a function of wavelength, as expected
due to the different temperatures of starspots and the surrounding
photosphere. The multi-colour data allowed a precise measurement
of the temperature of each starspot. The values that we found are
well in agreement with those found for other dwarf stars.

($iv$) The starspots detected in the GROND transits \#1 and \#2
present similar characteristics in terms of size, temperature and
latitude, suggesting that we observed the same starspot in two
transit events spaced by a time span during which Qatar-2\,A
performed one complete revolution. This allows a precise
measurement of the rotation period of Qatar-2\,A,
$P_{\mathrm{rot}}=11.4 \pm 0.5\,\mathrm{d}$ at a colatitude of
$72^{\circ}$, and the sky-projected spin orbit alignment,
$\lambda=4.3^{\circ} \pm 4.5^{\circ}$. The latter result implies
that the orbital plane of Qatar-2\,b is well aligned with the
rotational axis of its parent star.

($v$) Thanks to the ability of GROND and BUSCA to measure stellar
flux simultaneously through different filters, covering quite a
large range of optical wavelengths, we were able to search for a
radius variation of Qatar-2\,b as a function of wavelength. All of
the measurements are consistent with a larger value of the
planet's radius in the $g$-band when compared with the redder
bands. This phenomenon is attributable to unocculted starspots
which affect more strongly our measurements in the $g$ band. By
focussing on the results coming from the three close transits
observed with GROND, we reconstructed a more accurate transmission
spectrum of the planet's atmosphere in terms of the planet/star
radius ratio and compared it with three synthetic spectra, based
on model atmospheres in chemical equilibrium in which the presence
of strong absorbers were excluded, but with differing amounts of
Rayleigh scattering (Fig.\,\ref{Fig:12}). If we correct the
$g^{\prime}$-band by the amount indicated by atmospheric models,
the comparison between experimental data and synthetic spectra
suggests that the atmosphere of Qatar-2\,b could be dominated by
Rayleigh scattering at bluer wavelengths. The low value of the
radius observed between $700$ and $800$\,nm should be explicable
by a lack of potassium in the atmosphere of the planet (probably
caused by star-planet photoionization processes). This hypothesis
could be investigated with narrower-band filters.

\begin{figure*}
\centering
\includegraphics[width=18.cm]{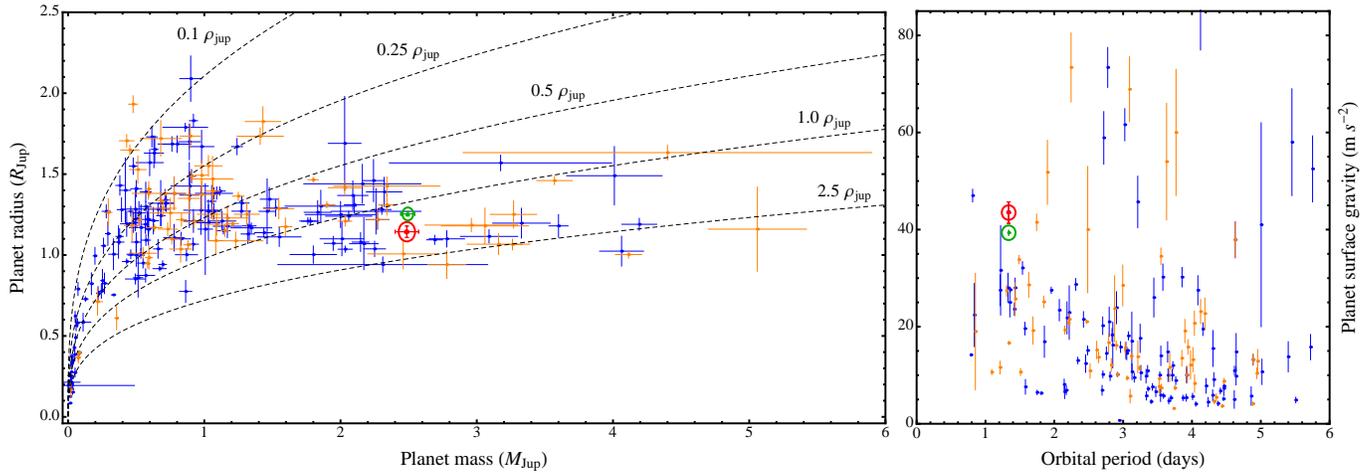}
\caption{\emph{Left-hand panel}: plot of the masses and radii of
the known TEPs. The orange symbols denote values from the
Homogeneous Studies project and the blue symbols results for the
other known TEPs. Qatar-2\,b is shown in red \citep{bryan2012} and
green (this work). Dotted lines show where density is 2.5, 1.0,
0.5, 0.25 and 0.1 $\rho_{\mathrm{Jup}}$. \emph{Right-hand panel}:
plot of the orbital periods and surface gravities of the known
TEPs. The symbols are the same as for left-hand panel. Data taken
from Transiting Extrasolar Planet Catalogue, available at
http://www.astro.keele.ac.uk/jkt/tepcat/ \citep{southworth2011}.}
\label{Fig:13}
\end{figure*}


\section*{Acknowledgements}
This paper is based on observations collected with the
MPG/ESO\,2.2-m located at ESO Observatory in La Silla, Chile; CAHA
2.2-m and 1.23-m telescopes at the Centro Astron\'{o}mico Hispano
Alem\'{a}n (CAHA) at Calar Alto, Spain; Cassini\,1.52-m telescope
at the OAB Loiano Observatory, Italy. Supplementary data were
obtained at the Canis-Major Observatory in Italy. Operation of the
MPG/ESO\,2.2-m telescope is jointly performed by the Max Planck
Gesellschaft and the European Southern Observatory. Operations at
the Calar Alto telescopes are jointly performed by the Max-Planck
Institut f\"{u}r Astronomie (MPIA) and the Instituto de
Astrof\'{i}sica de Andaluc\'{i}a (CSIC). GROND was built by the
high-energy group of MPE in collaboration with the LSW Tautenburg
and ESO, and is operated as a PI-instrument at the MPG/ESO\,2.\,2m
telescope. We thank Timo Anguita and R\'egis Lachaume for
technical assistance during the GROND observations. We thank David
Galad\'{i}-Enr\'{i}quez and Roberto Gualandi for their technical
assistance at the CA 1.23\,m telescope and Cassini telescope,
respectively. L.M.\ thanks Antonino Lanza for useful discussion.
J.S.\ acknowledges financial support from STFC in the form of an
Advanced Fellowship. The reduced light curves presented in this
work will be made available at the CDS
(http://cdsweb.u-strasbg.fr/). We thank the anonymous referee for
their useful criticisms and suggestions that helped us to improve
the quality of the present paper. The following internet-based
resources were used in research for this paper: the ESO Digitized
Sky Survey; the NASA Astrophysics Data System; the SIMBAD data
base operated at CDS, Strasbourg, France; and the arXiv scientific
paper preprint service operated by Cornell University.

%

\appendix

\section{S/N estimations}
\label{Appendix_A}

\begin{figure*}%
\centering
\includegraphics[width=12.cm]{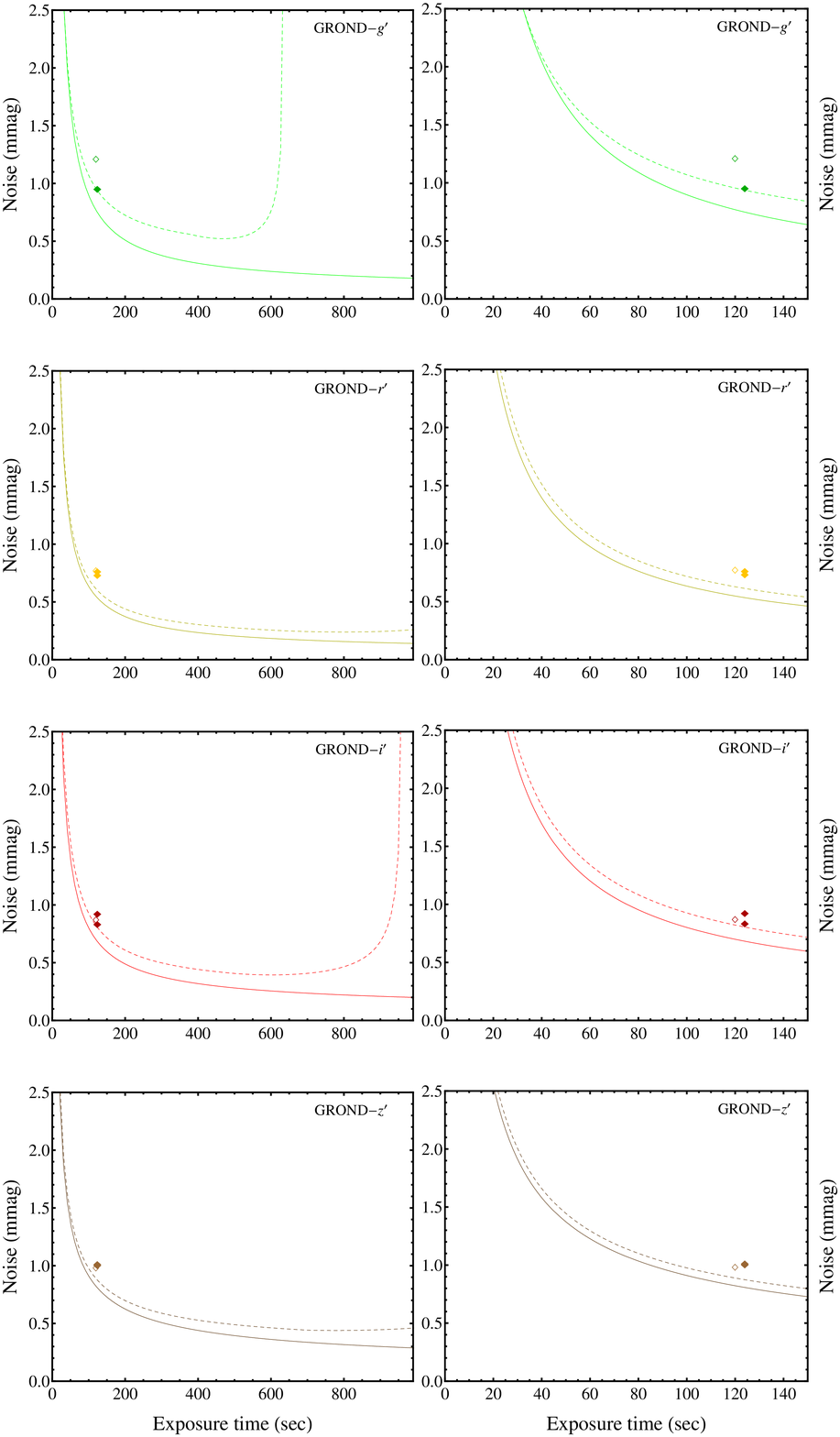}
\caption{Predicted noise levels for the GROND observations
presented in this work, but as a function of exposure time. The
solid curves show the predicted noise level per observation for
dark time and high observational cadence, whereas the dashed
curves that per observation in bright time and low cadence. Panels
in the left-hand column are for each of the four GROND optical
bands. Panels in the right-hand column are the same of left-hand
column, but zoomed to $t_{\mathrm{exp}}<150$. Diamonds represent
the measurements of the two Qatar-2\,b transits in dark time with
a cadence of 150 s and empty diamonds are for the measurements of
the transit in bright time with a cadence of 200 s (cfr. Table
\ref{Table:1}).}
\label{Fig:A01}
\end{figure*}

In order to test the goodness of the measurements reported in this
paper, we present signal-to-noise ratio (S/N) expectations for the
simultaneous multi-band photometric observations of Qatar-2 with
GROND. For each of the four GROND optical bands, we can quantify
the ratio of noise-to-signal per unit time by
\begin{equation}
\rho =
\sigma_{\mathrm{total}}\sqrt{t_{\mathrm{exp}}+d_{\mathrm{readout}}},
\end{equation}
where $t_{\mathrm{exp}}$ is the exposure time and
$d_{\mathrm{readout}}$ is the total dead time per observation --
the latter quantity is generally dominated by the CCD readout
time, but for GROND we have also to consider an extra dead time
due to the synchronization of the optical observations with the
NIR ones. $\sigma_{\mathrm{total}}$ is the total noise in a
specific band for 1 mag measurement in one observation and takes
into account five noise contributions that are added in
quadrature. They are the Poisson noise from the target and
background, readout noise, flat-fielding noise and scintillation,
i.e.
\begin{equation}
\sigma_{\mathrm{total}} =
\sqrt{\sigma_{\mathrm{target}}^{2}+\sigma_{\mathrm{sky}}^{2}+\sigma_{\mathrm{ron}}^{2}+\sigma_{\mathrm{flat}}^{2}+\sigma_{\mathrm{scint}}^{2}}.
\end{equation}
Following the procedure described by \citet{southworth2009al1}, we
performed S/N calculations for the GROND observations presented in
this work. The count rates for the target (overall) and for the
sky background (per pixel) were gathered from the {\sc
SIGNAL}\footnote{Information on the Isaac Newton Group's {\sc
SIGNAL} code can be found at http://catserver.ing.iac.es/signal/.}
code, scaling for the difference in telescope aperture and CCD
plate scale. The magnitudes of Qatar-2 in Sloan $g$, $r$ and $i$
were taken from AAVSO\footnote{The American Association of
Variable Star Observers (AAVSO) is a non-profit worldwide
scientific and educational organization of amateur and
professional astronomers.} archive. The magnitude in Sloan $z$ was
obtained by interpolating the previous ones with $J$, $H$ and $K$
magnitudes taken from the NOMAD catalog \citep{zacharias2004}.
Other input parameters, specific for each band, were the readout
noise of the CCD detector, the signal per pixel from the target
averaged over the PSF, the maximum total count in a pixel (target
plus sky) and the number of pixels inside the annulus of the
target to apply flat-field noise. The resulting curves of noise
level per observation are plotted in millimagnitudes as a function
of $t_{\mathrm{exp}}$ in Fig. \ref{Fig:A01} for dark time and high
observational cadence (solid curves) and for bright time and low
cadence (dashed curves), cfr. Table \ref{Table:1}.
The scatter of our measurements are a bit higher than the expected
noise, but fully consistent.

\end{document}